\documentclass{aa}  

\usepackage{graphicx}
\usepackage{txfonts}
\usepackage{natbib}
\bibpunct{(}{)}{;}{a}{}{,}
\begin{document}

   \title{A deep X-ray view of the bare AGN 
     Ark120.}

   \subtitle{IV. XMM-Newton and NuSTAR spectra dominated by two
     temperature (warm, hot) Comptonization processes.}

   \author{D.\ Porquet \inst{1}   
          \and
          J.~N.\ Reeves\inst{2,3} 
          \and
           G.\ Matt\inst{4}
          \and       
           A.\ Marinucci\inst{4}
          \and
           E.\ Nardini\inst{5}
          \and 
           V.\ Braito\inst{6}
          \and
           A.\ Lobban\inst{2}
           \and
           D.\ R.\ Ballantyne\inst{7} 
           \and \\
           S.~E.\ Boggs\inst{8}
           \and 
           F.~E.\ Christensen \inst{9}
           \and
           T.\ Dauser\inst{10}
           \and 
           D.\ Farrah\inst{11} 
           \and
           J.~ Garcia\inst{12,10,13}
           \and
           C.~J.\ Hailey\inst{14}
           \and
           F.\ Harrison\inst{12}
           \and
           D.\ Stern\inst{15}
          \and \\
           A.\ Tortosa\inst{4}
          \and 
           F.\ Ursini\inst{16}
          \and
          W.~W.\ Zhang\inst{17}
 }

   \institute{Universit{\'e} de Strasbourg, CNRS, Observatoire astronomique de Strasbourg, UMR 7550, F-67000 Strasbourg, France\\
              \email{delphine.porquet@astro.unistra.fr}
         \and Astrophysics Group, School of Physical \& Geographical
         Sciences, Keele University, Keele ST5 5BG, UK
         \and CSST, University of Maryland Baltimore County, 1000
         Hilltop Circle, Baltimore, MD 21250, USA
        \and Dipartimento di Matematica e Fisica, Universit{\`a} degli
        Studi Roma Tre, via della Vasca Navale 84, 00146, Roma, Italy
        \and INAF-Osservatorio Astrofisico di Arcetri, Largo Enrico
        Fermi 5, 50125, Firenze, Italy 
        \and INAF-Osservatorio Astronomico di Brera, via Bianchi 46,
        23807, Merate (LC), Italy
         \and Center for Relativistic Astrophysics, School of Physics,
         Georgia Institute of Technology, 837 State Street, Atlanta,
         GA 30332-0430, USA 
         \and Space Sciences Laboratory, University of California,
         Berkeley, CA 94720, USA
         \and 
        DTU Space, National Space Institute, Technical University of
        Denmark, Elektrovej 327, DK-2800 Lyngby, Denmark 
        \and Dr Karl Remeis-Observatory and Erlangen Centre for
        Astroparticle Physics, Sternwartstr. 7, D-96049 Bamberg,
        Germany
        \and 
        Department of Physics, Virginia Tech, Blacksburg, VA 24061, USA
       \and Cahill Center for Astronomy and Astrophysics, California
          Institute of Technology, Pasadena, CA 91125, USA 
        \and Harvard-Smithsonian Center for Astrophysics, 60 Garden
        Street, Cambridge, MA 02138, USA
        \and Columbia Astrophysics Laboratory, Columbia University,
        New York, New York 10027, USA
         \and Jet Propulsion Laboratory, California Institute of
         Technology, Pasadena, CA 91109, USA 
        \and INAF-IASF Bologna, Via Gobetti 101, I-40129 Bologna, Italy
        \and NASA Goddard Space Flight Center, Code 662, Greenbelt, MD
        20771, USA
             }

   \date{Received , 2017; accepted , 2017}

 
  \abstract
{}
   {[Abstract Truncated] We perform an X-ray spectral analysis of
     the brightest and cleanest bare AGN known so far, Ark\,120, in order to determine the process(es) at work in the
     vicinity of the SMBH.}
   {We present spectral analysis of data from an extensive 
    campaign observing Ark\,120 in X-rays 
     with {\sl XMM-Newton} (4$\times$120\,ks, 2014 March 18--24), and {\sl
       NuSTAR} (65.5\,ks, 2014 March 22).}
   {During this very deep X-ray campaign, the source was caught in a
     high flux state similar to the earlier
2003 {\sl XMM-Newton} observation, and about twice as bright as the
lower-flux observation in 2013. The spectral analysis confirms the
``softer when brighter'' behaviour of Ark\,120. 
The four {\sl XMM-Newton}/pn spectra are characterized by the presence of a prominent
  soft X-ray excess and a significant Fe\,K$\alpha$ complex. The continuum is
  very similar above about 3\,keV, while significant variability is present for the soft X-ray
excess. We find that relativistic reflection from
a constant-density, flat accretion disk cannot
simultaneously produce the soft excess, broad Fe\,K$\alpha$ complex, and hard X-ray excess. 
Instead, Comptonization reproduces the broadband (0.3--79 keV) continuum well, together with a contribution from a mildly relativistic disk reflection spectrum.}   
   {During this 2014 observational campaign, the soft X-ray spectrum of Ark
120 below $\sim$0.5\,keV was found to be dominated by Comptonization of seed photons from
the disk by a warm ($kT_{\rm e}$$\sim$0.5\,keV), optically-thick
corona ($\tau$$\sim$9). 
Above this energy, the X-ray spectrum becomes dominated by Comptonization from electrons in
a hot optically thin corona, while the broad Fe\,K$\alpha$ line and
the mild Compton hump result from reflection off
the disk at several tens of gravitational radii.}

   \maketitle
%

\section{Introduction}

In the standard picture, the emission of an active galactic nucleus
(AGN) stems from an accretion disk around a supermassive black hole (SMBH) with
mass spanning from a few millions to billions of solar masses.
X-ray spectra offer a unique potential to probe matter very close
to the black hole and to measure the black hole spin. 
The X-ray spectra of AGN usually exhibit one
or more of the following components: a soft
excess below 2 keV, a power-law continuum up to about 10\,keV, 
a Fe\,K${\alpha}$ line complex near 6.4 keV, and a Compton scattering hump 
near 20--30\,keV. 

Recent studies using {\sl XMM-Newton} have shown that the soft
X-ray excess component is commonly seen in AGN,   
and that for most AGN (the exception possibly being low-mass
Narrow Line Seyfert 1 galaxies; \citealt{Done12}) this soft excess is
not the hard tail of the big blue bump 
observed in the UV coming from blackbody emission of the
accretion disk \citep[e.g.,][]{GD04,P04a,Piconcelli05}. 
Different explanations have been proposed: 
e.g., photo-ionized emission blurred by relativistic
motion in the accretion disk \citep[e.g.,][]{Crummy06}; Comptonization
of soft (extreme UV) seed photons by the electrons of a corona above
the disk \citep[e.g.,][]{Czerny87}; and an artifact of
strong, relativistically smeared, partially 
ionized absorption \citep{GD04}. 
These models can give a good representation of the
soft excess, though the current simulations of line-driven AGN accretion disk winds
cannot reproduce the smooth, soft X-ray excess \citep{Schurch09}. Therefore,
the {\it origin of the soft excess is still an 
open issue}. 

Besides, the study of the Fe\,K$\alpha$ complex emission can be used 
to provide fundamental diagnostics of the physical
and dynamical conditions of the AGN central engine from the inner part of
the accretion disk to the far-away molecular torus.
Relativistic (or broad) Fe\,K$\alpha$ lines have been studied 
in numerous AGN \cite[e.g.,][]{Nandra07,Brenneman09,Patrick12};
especially \object{MCG-06-30-15}
\citep[e.g.,][]{Tanaka95,Fabian02,Marinucci14}, and 
other particular AGN thanks notably to {\sl XMM-Newton} and/or {\sl NuSTAR} data 
(\object{NGC 1365}:  
\citealt{Risaliti13} and \citealt{Walton14}; \object{Mrk\,335}: \citealt{Parker14};
 and \object{SWIFT J2127.4+5654}: \citealt{Marinucci14}). 
Yet, the physical interpretation of the observed broad
Fe\,K$\alpha$ lines has been disputed due to the common presence of a
warm absorber \citep{P04a,Piconcelli05,Blustin05}, 
which has been proposed to explain in part the broadness of the lines 
\citep{Turner09}. Indeed, the presence of this hot gas on the line-of-sight could 
severely complexify the X-ray data analysis,   
in particular by distorting  the underlying continuum of the Fe\,K line. 
However, arguments against this
interpretation have been invoked by \cite{Reynolds09b}. 

Likewise, several origins for the hard X-ray spectrum above 10\,keV 
for type 1 Seyferts have been proposed, such
as relativistic reflection, complex absorption, and Comptonization, or
a combination thereof 
\citep[e.g.,][]{Nardini11,Noda11,Patrick11,Patrick12,Risaliti13,Walton14,Ursini15,Mehdipour15}.
Therefore, AGN showing no (or very weak) presence of any X-ray
warm absorber -- so-called {\it ``bare AGN''} -- are the best
targets to directly investigate the process(es) at work in the vicinity
of SMBHs. \\

 Ark\,120 (z=0.033, $M_{\rm
  BH}$=1.50$\pm$0.19$\times$10$^{8}$\,M$_{\odot}$\footnote{Black hole mass
  determined via reverberation mapping \citep{Peterson04}.}) 
is the brightest and cleanest
bare AGN known so far, i.e. displaying neither intrinsic reddening in
its IR continuum  nor evidence for absorption in UV and X-rays 
\citep[e.g.,][]{Ward87,Crenshaw99,Reynolds97}. 
The first {\sl XMM-Newton} observation performed in August 2003 ($\sim$80\,ks pn net exposure time) confirmed that its spectrum is warm
absorption-free, with the smallest upper limit to the column density found among AGN
($\sim$3$\times$10$^{19}$\,cm$^{-2}$; \citealt{Vaughan04}). 
Ark\,120 therefore 
represents the best target to have the ``purest'' view of the
  properties of the accretion disk and of the black hole spin in AGN. 
This object displays a prominent soft
excess observed down to 0.3\,keV and a significant 
Fe\,K$\alpha$ line complex \citep{Vaughan04}.  
The width of the broad component of the Fe\,K$\alpha$ line
(FWHM$~\sim$30\,000\,km\,s$^{-1}$) was much larger than that of the broad
optical lines from the broad line region (BLR),  
with FWHM(H$_{\beta}$)=5\,850$\pm$480\,km\,s$^{-1}$ (\citealt{Wandel99b}).
Ark\,120 was 
also observed with {\sl Suzaku} in April 2007 ($\sim$100\,ks) confirming the
presence of a large soft excess and of a significant broad Fe\,K$\alpha$ line 
\citep{Nardini11,Patrick11}. 
Recently, \cite{Matt14} reported the first spectral analysis of a
simultaneous {\sl XMM-Newton} and {\sl NuSTAR} observation performed
in February 2013. They found that the smooth soft excess was more
likely explained by Comptonization. 
In contrast with the 2003 {\sl XMM-Newton} observation
\citep{Vaughan04} and the 2007 {\sl Suzaku} observation
\citep{Nardini11}, while a significant soft excess was present too, 
no obvious signature for relativistic
reflection was found in the 2013 observation. It is worth 
mentioning that in 2013 the flux of Ark\,120 was lower by about 
a factor of two than during the 2003 observation. 
Both this lower flux and the lack of any relativistic signature in this 
2013 observation (``low-flux state'') may be explained by the presence
of an extended 
optically-thick corona which hides most of the relativitic reflection from the
 accretion disk, while in 2003 (``high-flux state'' ) and in
2007 (``intermediate-flux state'') this corona was likely
less thick and/or less extended \citep{Matt14}. 

An extensive X-ray observational campaign
was performed from 2014 March 18 to March 24 to study Ark\,120 in order 
to directly probe the accretion disk properties and the SMBH 
spin in this moderate Eddington ratio AGN
(\.{M}$\gtrsim$0.05\,\.{M}$_{\rm Edd}$; \citealt{Vaughan04}).
This campaign combined a deep {\it XMM-Newton}
observation (480\,ks  split into
  four consecutive 120\,ks observations from 2014 March 18 to March
  24; PI: D.\ Porquet) 
with a simultaneous 120\,ks {\it Chandra}/HETG
observation\footnote{This was the first {\it
  Chandra} observation of Ark\,120. The observation was split in three
 consecutive sequences as described in Reeves et al.\ (2016) and
Nardini et al.\ (2016).} (PI: D. Porquet). Furthermore, a {\sl
NuSTAR} observation (65\,ks; PI: {\sl NuSTAR} AGN team)  
was performed during the third {\sl XMM-Newton} 
observation, i.e., on 2014 March 22. 

In \cite{Reeves16} (hereafter Paper\,I), we
reported on the analysis of the soft X-ray spectrum using the 
480\,ks {\sl XMM-Newton}/RGS and 120\,ks {\sl Chandra}/HETG spectra. 
We confirmed that there were no detectable absorption lines due to the
warm absorber in the deep RGS spectra, 
and that Ark\,120 is the cleanest bare nucleus AGN known so far. 
Only absorption lines from the interstellar medium of our Galaxy were found. 
Interestingly, several soft X-ray emission lines  from the He-like and
H-like ions of N, O, Ne and Mg were revealed for the first time 
thanks to this very high signal-to-noise (S/N) RGS spectrum. 
As a consequence, Ark\,120 is not intrinsically bare since substantial X-ray emitting
gas is present out of the direct line of sight towards this AGN.
This result is very important in the framework of the unified scheme
of AGN, which invokes the existence
of wide scale obscuring and emitting gas \citep{Antonucci93}. 

In \cite{Nardini16} (hereafter Paper II) 
we took advantage of the unprecedented depth of the new data sets to 
study the properties of the composite emission complex due to iron fluorescence at 
6--7 keV. The most prominent feature peaks around 6.4 keV, and can be plainly 
identified with the K$\alpha$ transition from neutral iron. The
profile of the narrow Fe\,K$\alpha$ core is resolved in the {\sl
  Chandra}/HETG spectrum with a FWHM of 4\,700$^{+2700}_{-1500}$\,km\,s$^{-1}$, consistent with a BLR origin (as was also found for the soft X-ray emission 
lines; Paper\,I). Excess components are systematically detected redwards (6.0--6.3 keV) and 
bluewards (6.5--7.0 keV) of the narrow Fe\,K$\alpha$ core. The energy and equivalent width 
of the red wing rule out an interpretation in the form of the Compton
shoulder of the 6.4-keV  
K$\alpha$ feature. Moreover, its variability over timescales of about
one year (February 2013 to March 2014) hints at the presence of an
emission component from the  
accretion disk. Excess emission maps and time-resolved spectra based
on the four consecutive orbits  
of \textit{XMM-Newton} monitoring show that both the red and blue features are 
highly variable on short timescales (30--50 ks) but appear to be
disconnected.   
Such a timescale suggests an origin for these two components at a few
tens of gravitational radii from the central supermassive black hole, 
 potentially from discrete hot spots on the disk surface.  

In \cite{Lobban17} (submitted; hereafter Paper\,III), we presented the
spectral/timing properties of Ark\,120 using all available {\it
  XMM-Newton} data (including this XMM-Newton Large Programme), 
a recent $\sim$6-month {\it Swift} monitoring
 campaign \citep{Gliozzi17}, and data from {\sl RXTE} obtained between
 1998 and 2006.   
The spectral decomposition was investigated through
fractional rms, covariance and difference spectra, where we found that
the mid- to long-timescale ($\sim$day-year) variability is dominated
by a relatively smooth, steep component, which peaks in the soft X-ray
band.  Additionally, we found evidence for a variable component of
Fe\,K$\alpha$ emission on the red side of the near-neutral Fe\,K$\alpha$ core
on long timescales, consistent with Paper\,II.  We also
measured the power spectrum and searched for frequency-dependent
Fourier lags, obtaining the first detection of a high-frequency soft
X-ray lag in this source. Finally, we found well-correlated
optical/UV/X-ray variations with the {\it Swift} UVOT and searched for
multi-wavelength time delays, finding evidence for the optical emission
lagging behind the X-rays with a time delay of $\tau = 2.4 \pm1.8$\,days.

In this Paper IV, we report on the X-ray spectral analysis 
of the four 120\,ks {\sl XMM-Newton}/pn time-averaged spectra 
 performed in March 2014, 
which represent the deepest and longest elapsed time X-ray observation for a bare AGN.  
We also report on a {\sl NuSTAR} observation that was simultaneous
with the third 2014 {\sl XMM-Newton} observation (see Table~\ref{tab:log} for
details). 
In section~\ref{sec:obs}, we describe the observations, the
 data reduction, and the spectral analysis method. 
The spectral analysis of the four 2014 {\sl XMM-Newton}/pn spectra is presented in
section~\ref{sec:2014allpn}, 
and the broad-band X-ray spectrum ({\sl XMM-Newton} and {\sl NuSTAR})
of the 2014 March 22 observation in section~\ref{sec:broadbandX}. 
In section~\ref{sec:summary} our main results are summarized, before
the discussion and conclusions in section~\ref{sec:discussion}.

\section{Observations,  data reduction and analysis}\label{sec:obs}

\subsection{XMM-Newton and NuSTAR data reduction}

\begin{table*}[t!]
\caption{Observation log of the data analysed in this work from the
  2014 Ark\,120 observational campaign.}
\label{to}
\begin{tabular}{c@{\hspace{15pt}}r@{\hspace{15pt}}c@{\hspace{15pt}}c@{\hspace{15pt}}c@{\hspace{15pt}}c}
\hline \hline
Mission & Obs.\,ID & Obs.\,Start (UTC) & Exp.$^a$ & $\mathcal{C}^b$
(s$^{-1}$) \\
\hline
{\sl XMM-Newton} & 0721600201 & 2014 March 18 -- 08:52:49 & 81.6 &
27.14$\pm$0.02 \\
{\sl XMM-Newton} & 0721600301 & 2014 March 20 -- 08:58:47 & 83.9 &
22.65$\pm$0.02 \\
{\sl XMM-Newton} & 0721600401 & 2014 March 22 -- 08:25:17 & 82.4 &
25.23$\pm$0.02 \\
{\sl NuSTAR}     & 60001044004 & 2014 March 22 -- 09:31:07 &65.5 &
1.089$\pm$0.004 (FPMA) \\
          &   &  & 65.3& 1.072$\pm$0.004 (FPMB)\\
{\sl XMM-Newton} & 0721600501 & 2014 March 24 -- 08:17:19 & 81.9 & 22.78$\pm$0.02\\
\hline
\hline
\end{tabular}
\label{tab:log}
\flushleft
\small{\textit{Notes.} $^a$Net exposure in ks. $^b$Source count rate
	over 0.3---10 keV for {\sl XMM-Newton}/pn and over 3--79\,keV for {\sl NuSTAR}.} 
\end{table*}

During this observational campaign Ark 120 was observed by {\sl
  XMM-Newton} over four consecutive orbits between 2014 March 18 and March
24  (Table~\ref{tab:log}). As reported in Paper\,II, the event files were
reprocessed with the Science Analysis System (SAS) v14.0, applying the
latest calibrations available in 2015 February. Due to the high source
brightness, the EPIC instruments were operated in Small Window mode. 
However, this observation mode was not sufficient to prevent pile-up in
the MOS cameras, and therefore only the EPIC/pn \citep{Struder01} 
data are taken into account (selecting the event patterns 0-4, i.e.,
single and double pixels). The four pn spectra were extracted 
from a circular region centered on Ark\,120, with a radius of
30${\arcsec}$ to avoid the edge of the chip. The background
spectra were extracted from a rectangular region in the lower part of
the small window that contains no (or negligible) source photon.  
 The latest part of each orbit was not used due to high background
  flaring level.   
After the correction for dead time and background flaring, the total net
exposure was about 330\,ks. Redistribution matrices and ancillary
response files for the four pn spectra were generated with the SAS tasks
{\sc rmfgen} and {\sc arfgen}.  
As shown in Paper\,III, there is a significant flux
variability during each of the four {\sl XMM-Newton} observations and between them. 
 However, the spectral variability within any single orbit is slow and moderate, so we are
 able to use time-averaged spectra for each of the four observations. 
As detailed in Paper\,II, a gain shift has to be
applied to take into account the known inaccuracy of the EPIC/pn
energy scale likely due to inaccuracies in the long-term charge
transfer (CTI) calibration\footnote{See
  http://xmm2.esac.esa.int/docs/documents/CAL-SRN-0300-1-0.pdf and \cite{Marinucci14}.}. The
corresponding values for the {\sc xspec} {\sl gain} function are
reported in Paper\,II. The 0.3--10\,keV pn spectra were
binned to give 50 counts per bin.

{\sl NuSTAR} \citep{Harrison13} observed Ark 120 with its two
co-aligned X-ray telescopes with corresponding Focal Plane Modules A 
(FPMA) and B (FPMB) starting on 2014 March 22 for a
total of $\sim$131\,ks of elapsed time.  The Level 1 data products were
processed with the {\sl NuSTAR} Data Analysis Software (NuSTARDAS)
package (v.~1.6.0). Cleaned event files (level 2 data products) were
produced and calibrated using standard filtering criteria with the
\textsc{nupipeline} task and the calibration files available in the
{\sl NuSTAR} calibration database (CALDB: 20170222). Extraction
radii for both the source and the background spectra were $1.25$
arcmin. After this process, the net exposure time for the observation
was about 65\,ks, with most of the time lost to Earth occultations.
The pair of 3.5--79\,keV {\sl NuSTAR} spectra were binned in order to
over-sample the instrumental resolution by at least a factor of 2.5
and to have a S/N ratio greater than 5 in
each spectral channel.

The summary log of the Ark\,120 X-ray observations used in this work from
the 2014 observational campaign are reported in
Table~\ref{tab:log}.

\subsection{Spectral analysis method}

The {\sc xspec} v12.9.0g software package \citep{Arnaud96}  
was used for the spectral analysis.
 The Galactic column density is assumed to be $N_{\rm H}$=9.78$\times$10$^{20}$\,cm$^{-2}$ 
as inferred from the weighted average $N_{\rm H}$ value of the LAB Survey of
Galactic \ion{H}{i} \citep{Kalberla05}.  
Since there can be some additional contribution associated with
molecular hydrogen \citep{Willingale13}, we allow the value of
Galactic $N_{\rm H}$ to vary slightly (except for the fits above 3\,keV, where
the value is fixed to 9.78$\times$10$^{20}$\,cm$^{-2}$). 
However, we do not allow for any intrinsic absorption in the rest frame of 
Ark\,120, since, as found in Paper\,I from the deep RGS spectrum, none
is observed. 
We used the X-ray absorption model {\sc tbnew (v2.3.2)} from
\cite{Wilms00}, assuming throughout their interstellar medium (ISM) elemental abundances and
the cross-sections from \cite{Verner96}. 

As reported in Paper\,II, the narrow neutral core of the
Fe\,K$\alpha$ emission complex is consistent with being associated
with the BLR, and so is some 
contribution to the H-like line of iron. Consequently, throughout this work 
we take into account the contribution from the BLR
to the Fe\,K complex using three Gaussian lines:
 the Fe\,K$\alpha_{\rm BLR}$ ($E$ fixed at 6.40\,keV) plus its
 associated Fe\,K$\beta_{\rm BLR}$ 
 line ($E$ fixed at 7.05\,keV), and the H-like iron line ($E$ fixed at
 6.97\,keV).  
 The normalization of Fe\,K$\beta_{\rm BLR}$ is set to 0.135 times
 that of Fe\,K$\alpha_{\rm BLR}$ \citep{Palmeri03}.   
The widths of these three lines are fixed to the value inferred in
Paper\,II for the Fe\,K$\alpha$ narrow core, i.e., 43\,eV, as
determined from the simultaneous {\sl Chandra}/HETG spectrum (Paper\,II). 
These three BLR emission lines are called hereafter ``3 zgaussians(BLR)''.

Throughout this work, when considering relativistic reflection
modelling, we use the {\sc relxill} package 
(v0.4c\footnote{This package version is more recent than the one
used in Paper\,II (v0.4a), however the fit differences are
negligible.}; released in May 2016). 
These reflection models calculate the proper emission angle of
the radiation at each point on the accretion disk, and then take the
corresponding reflection spectrum into account. This angle-dependent
 model connects self-consistently the \linebreak {\sc xillver} 
\citep{Garcia10,Garcia11,Garcia13,Garcia14}
reflection models with the relativistic blurring code {\sc relline}
\citep{Dauser10,Dauser13,Dauser14,Dauser16}. 
In this version 0.4c the cosmological redshift is properly taken into account 
in the cut-off energy as well. 
The models used in this work assume a constant density profile of the accretion disk
with $n_{\rm e}$=10$^{15}$\,cm$^{-3}$ (except when mentioned
otherwise, see $\S$\ref{sec:nustarall}), and an exponentially
broken power-law as the intrinsic continuum shape. 
Two main geometries are possible: the coronal one ({\sc relxill}) and the
lamppost one ({\sc relxilllp}). Detailed descriptions of these
models and their corresponding parameters are reported in Appendix~\ref{ref:relxill}. 

We use $\chi^{2}$ minimization throughout, quoting 90 percent errors
for one interesting parameter ($\Delta\chi^{2}$=2.71) unless otherwise
stated. Default values of H$_{\rm 0}$=70\,km\,s$^{-1}$\,Mpc$^{-1}$,
$\Omega_{\rm m}$=0.27, and $\Omega_{\Lambda}=0.73$ are assumed. 
All figures are displayed in the AGN rest-frame.

\section{Spectral analysis of the four 2014 XMM-Newton pn
  observations}\label{sec:2014allpn}

\begin{table}[t!]
	\caption{Simultaneous fit of the four 2014 { \sl XMM-Newton}/pn spectra with the baseline
 relativistic reflection model (model $\mathcal{A}$) over the 3--10\,keV energy range.  
Hydrogen column density is fixed to the Galactic value, 
i.e. 9.78$\times$10$^{20}$\,cm$^{-2}$. {\it nc} means that the
parameter value is not constrained. (a) the assumed value of the spin
has no impact on the other inferred parameters of the fit. (f) means
that the parameter is fixed.}             
\label{table:1}    
\centering                          
\begin{tabular}{l c c c}        
\hline\hline                 
Parameters & \multicolumn{1}{c}{relxill}& \multicolumn{1}{c}{relxill}&\multicolumn{1}{c}{relxilllp}\\    
\hline                       
\vspace{0.5mm}
$q$     &   $\leq$1.1   & 3(f)     & $-$   \\
\vspace{0.5mm}
$R_{\rm in}$ ($R_{\rm ISCO}$) &  1 (f)   & 17.8$^{+32.7}_{-8.6}$   & 1 (f) \\
\vspace{0.5mm}
$h$ ($R_{\rm g}$)&  $-$   &  $-$ & 93$^{+29}_{-25}$  \\
\vspace{0.5mm}
$a$         & {\it nc} & 0.0 (f)$^{(a)}$           & {\it nc}    \\
\vspace{0.5mm}
$\theta$ (degrees)      & 38.0$^{+9.5}_{-5.4}$  &  31.1$^{+6.5}_{-12.1}$       & 35.1$^{+3.9}_{-4.1}$   \\
\vspace{0.5mm}
log\,$\xi$ (erg\,cm\,s$^{-1}$)        & 2.4$\pm$0.1  & 2.4$\pm$0.1        & 2.6$\pm$0.1  \\
\vspace{0.5mm}
$A_{\rm Fe}$      & $\leq$0.6 & $\leq$0.6     & $\leq$0.7 \\
\hline                                  
                          &   \multicolumn{2}{c}{2014 March 18} \\
\hline                                  
\vspace{0.5mm}
$\Gamma$        & 1.92$^{+0.02}_{-0.03}$  & 1.92$\pm$0.02 & 1.90$\pm$0.01 \\
$\mathcal{R}$            & 0.5$^{+0.2}_{-0.1}$ & 0.5$\pm$0.1      & $-$ \\
\vspace{0.5mm}
$norm$ ($\times$10$^{-4}$)       &  2.3$\pm$0.1 &  2.3$\pm$0.1 &   2.4$\pm$0.1 \\
\hline                                  
                &   \multicolumn{2}{c}{2014 March 20} \\
\hline                                  
\vspace{0.5mm}
$\Gamma$      & 1.88$^{+0.02}_{-0.01}$  &1.88$^{+0.01}_{-0.02}$      & 1.85$\pm$0.01  \\
\vspace{0.5mm}
$\mathcal{R}$            & 0.5$\pm$0.1   & 0.5$\pm$0.1     & $-$  \\
\vspace{0.5mm}
$norm$  ($\times$10$^{-4}$) &  2.2$\pm$0.1 &  2.2$\pm$0.1  & 2.3$\pm$0.1\\
\hline                                 
                &   \multicolumn{2}{c}{2014 March 22} \\
\hline                                  
\vspace{0.5mm}
$\Gamma$       & 1.88$^{+0.04}_{-0.02}$ & 1.88$^{+0.02}_{-0.03}$    &  1.86$\pm$0.03\\
\vspace{0.5mm}
$\mathcal{R}$           & 0.5$\pm$0.1 & 0.5$\pm$0.1      & $-$ \\
\vspace{0.5mm}
$norm$  ($\times$10$^{-4}$)       & 2.5$\pm$0.1  & 2.5$\pm$0.1  & 2.6$\pm$0.1  \\
 \hline
                &   \multicolumn{2}{c}{2014 March 24} \\
\hline                                  
\vspace{0.5mm}
$\Gamma$         &1.85$^{+0.02}_{-0.01}$ &1.85$^{+0.01}_{-0.02}$   & 1.84$\pm$0.01\\
\vspace{0.5mm}
$\mathcal{R}$             & 0.4$\pm$0.1& 0.4$\pm$0.1 & $-$   \\
\vspace{0.5mm}
$norm$  ($\times$10$^{-4}$)          &2.5$\pm$0.1  &2.5$\pm$0.1  &  2.5$\pm$0.1 \\
 \hline
 \hline
$\chi^{2}$/d.o.f.  & 4461.7/4568 & 4459.4/4569 & 4493.1/4572\\
\vspace{0.5mm}
$\chi^{2}_{\rm red}$ & 0.98 & 0.98 & 0.98\\
\hline    \hline                  
\end{tabular}
\label{tab:pn3-10}
\end{table}

\begin{figure}[t!]
\begin{tabular}{cc}
\includegraphics[width=0.9\columnwidth,angle=0]{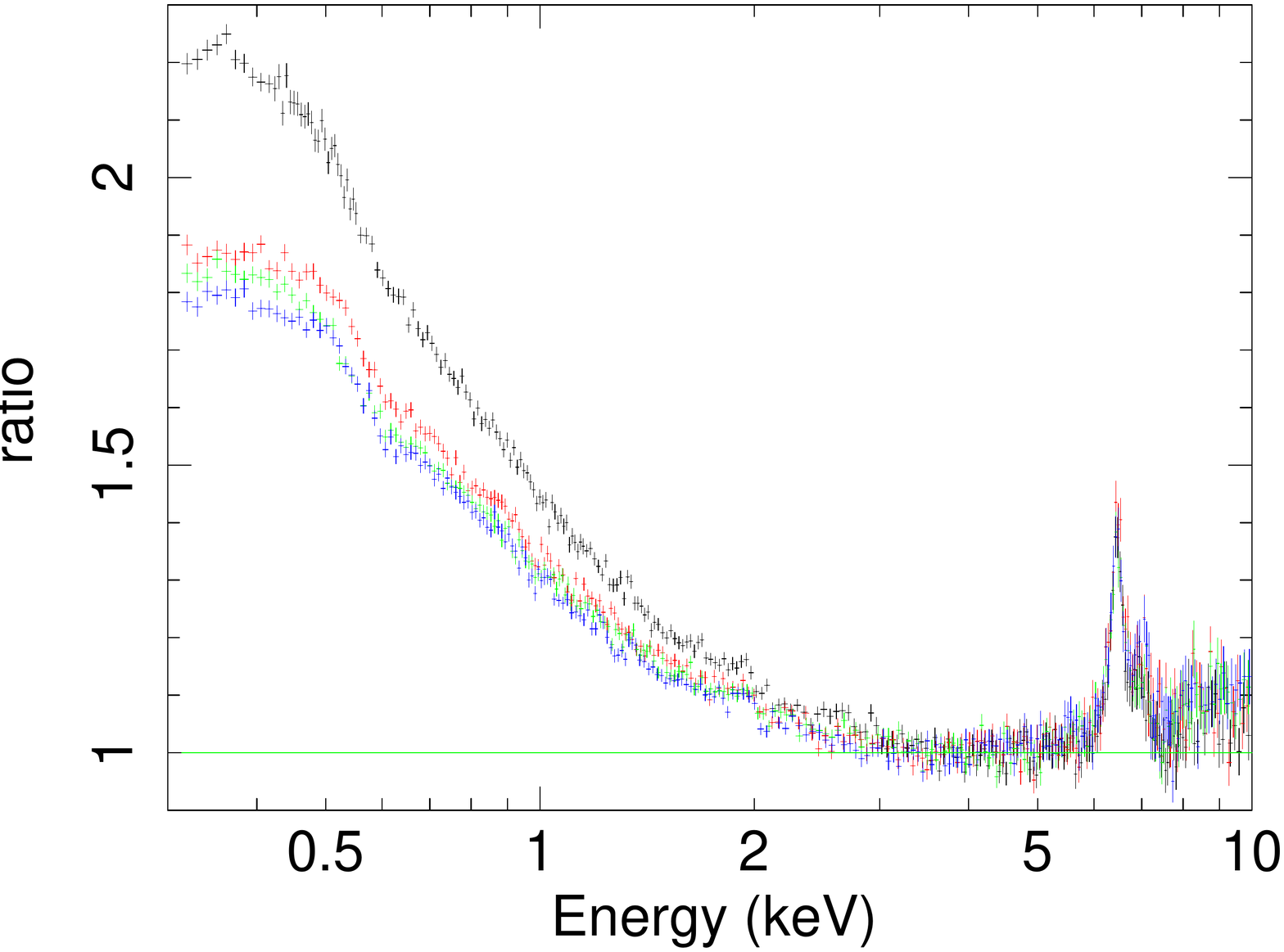}\\
\includegraphics[width=0.9\columnwidth,angle=0]{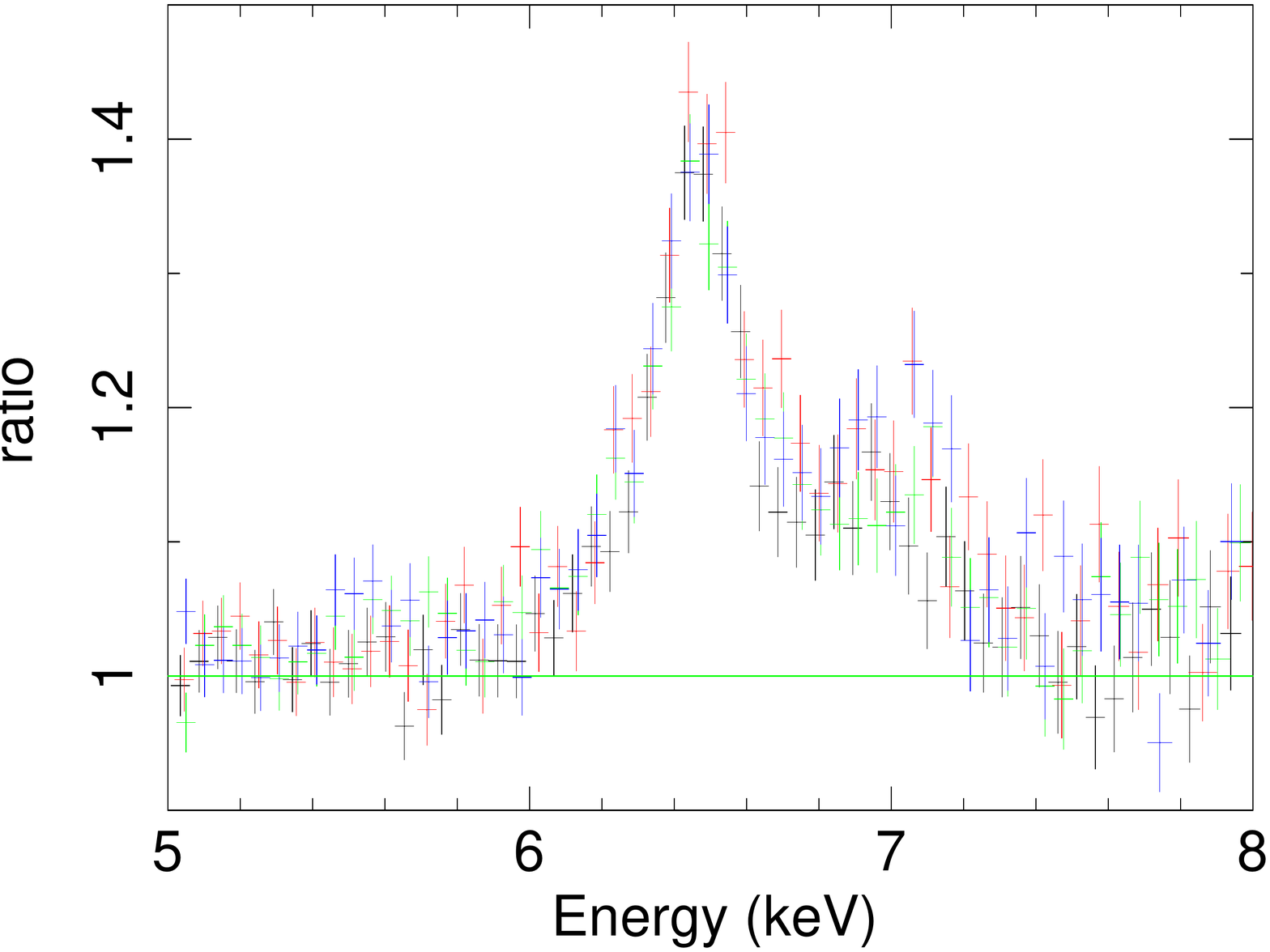}
\end{tabular}
	\caption{Data/model ratio of the four 2014 {\sl XMM-Newton}/pn spectra of Ark\,120 fitted 
 with a Galactic absorbed power-law continuum model over the
 3--5\,keV energy range, and then extrapolated over the 0.3--10\,keV
 energy range. Black: March 18, red: March 20,
 green: March 22, and blue: March 24. 
{\it Top panel}: 0.3--10\,keV energy range. 
{\it Bottom panel}: Zoom on the Fe\,K$\alpha$ complex. }  
\label{fig:3-5keV}
\end{figure}

In order to characterize the main X-ray components of the
spectra, we fit the four {\sl XMM-Newton}/pn spectra between 3--5\,keV using a simple
absorbed power-law model. The absorption column density has been fixed
to the Galactic one, i.e. 9.78$\times$10$^{20}$\,cm$^{-2}$. 
The power-law index is tied between the four spectra, while
the normalization is allowed to vary between the
observations ($\chi^{2}$/d.o.f.=1656.8/1591). 
We find a photon index of 1.87$\pm$0.02, which is typical of those
found for radio-quiet quasars \citep[e.g.,][]{P04a,Piconcelli05}. The
unabsorbed flux values between 3 and 5\,keV are about
1.8-2.0$\times$10$^{-11}$\,erg\,cm$^{-2}$\,s$^{-1}$ 
and are similar to that found for 
the 2003 {\sl XMM-Newton} observation, and about twice as bright as the 2013 observation.   
Then, we extrapolate over the whole 0.3--10\,keV energy range. 
As illustrated in Fig.~\ref{fig:3-5keV} (top panel), the four observations show a
significant 
soft X-ray excess below 2\,keV that is variable between the
observations (see also Paper\,III) with the first observation (2014 March 18)
exhibiting the largest soft X-ray excess.  
The prominent Fe\,K line
profile is consistently seen in each of the observations
(Fig.~\ref{fig:3-5keV}, bottom panel), though as
shown in Paper\,II the red and blue sides of the Fe\,K$\alpha$ complex
are variable on time-scale of about 10-15 hours, i.e faster than the
total duration of each observation.

\subsection{Spectral analysis above 3\,keV}

In this section, we aim to characterize the disk component(s) above
3\,keV, i.e., without any bias from the soft excess contribution.
This has been already investigated in Paper\,II, but here more general
relativistic reflection configurations are probed (e.g, non-solar iron
abundance, free inclination angle).

We use the baseline reflection model $\mathcal{A}$ defined as {\sc
  tbnew$\times$[relxill+3$\times$zgaussians(BLR)]}. 
 The photon index ($\Gamma$), the reflection fraction ($\mathcal{R}$), and the 
normalization ($norm$) of the underlying continuum of the 
relativistic reflection component 
are allowed to vary between each observation. 
The high-energy cut-off ($E_{\rm cut}$) is fixed to 1000\,keV since it
cannot be constrained from the pn energy range.

\begin{figure}[t!]
\begin{tabular}{c}
\includegraphics[width=1\columnwidth,angle=0]{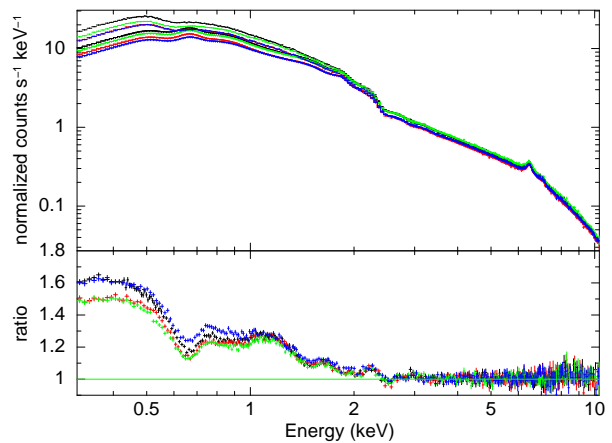}\\
\end{tabular}
\caption{Extrapolation down to 0.3\,keV of the fit over the 3--10\,keV energy range 
	of the four 2014 {\sl XMM-Newton}/pn spectra with the baseline relativistic reflection 
model (model $\mathcal{A}$), where $q$ is allowed to vary and $R_{\rm in}$ is fixed
to the ISCO. The fit parameters are reported in Table~\ref{tab:pn3-10} (column 2).  
Black: March 18, red: March 20,
 green: March 22, and blue: March 24. 
} 
\label{fig:3-10extra}
\end{figure}

We first consider a coronal geometry assuming a single 
power-law disk emissivity index $q$ ($\epsilon \propto R^{-q}$), tied 
between the four observations, and fixing the
inner radius of the reflection component ($R_{\rm in}$) at 
the innermost stable circular orbit (ISCO) which is self consistently
determined from the spin value in the relxill models. 
We find a good fit ($\chi^{2}/$d.o.f.=4461.7/4568) and infer a
very flat emissivity index of $\leq$1.1 (see Table~\ref{tab:pn3-10}). 
The photon indices ($\Gamma\sim$1.85--1.92) are typical of what is found in type-1 AGN 
\citep[e.g.,][]{P04a,Piconcelli05}. 
The extrapolation of the fit down to 
0.3\,keV shows that the soft excess is not accounted for by this
model, which leaves a large positive residual below 2\,keV (Fig.~\ref{fig:3-10extra}). 
If, instead, we fix the disk emissivity index to the standard value of
3, the spin value to 0,
and allow $R_{\rm in}$ (expressed in R$_{\rm ISCO}$ units) to
vary, we also find a good fit in the 3--10 keV  
energy range and derive $R_{\rm in}$= 17.8$^{+32.7}_{-8.6}$\,$R_{\rm
  ISCO}$ (see Table~\ref{tab:pn3-10}). If we fix the spin value to the maximal
ones, we find $R_{\rm in}$$\geq$56\,$R_{\rm
  ISCO}$ and $R_{\rm in}$=11.9$^{+22.3}_{-5.6}$\,$R_{\rm ISCO}$, for
a=0.998 and a=$-$0.998, respectively. 
This suggests that whatever the spin value is, the reflection 
component does not arise in the very 
inner part of the accretion disk, in agreement with the results
discussed in Paper\,II.  
The extrapolation of the fit down to 0.3\,keV shows that the soft
excess is not accounted for, with a similarly large positive residual. 
The reflection fraction is found to be rather low, with $\mathcal{R}$=0.4--0.5 
in both cases.  

Then we adopt a lamppost geometry\footnote{The ``lamppost'' model
  represents a X-ray source geometry in AGN, 
where the primary source of radiation is modelled by a
point source on the rotational axis of the black hole.}
 using the {\sc relxilllp} model.
The reflection fraction is calculated directly from the lamppost
geometry with the parameter {\tt fixReflFrac} fixed to 1. 
We find that the height of the X-ray source above the accretion disk
is rather high ($h$=93$^{+29}_{-25}$\,$R_{\rm g}$, see Table~\ref{tab:pn3-10}). 
This suggests that the disk illumination is not centrally
concentrated. As for the coronal geometry, a strong positive residual is found below
3\,keV when the best fit is extrapolated down to 0.3\,keV. 

To summarize, good fits are found using a combination of a primary
power law and a relativistic reflection component when
considering the 3--10\,keV energy range, 
but they indicate that reflection does not arise in the very inner part 
of the accretion disk (flat disk emissivity index, large $R_{\rm in}$ or $h$), 
and has a moderate reflection fraction of about 0.4--0.5 (coronal geometry).
These results are similar to those found in Paper\,II. 
Moreover, in all cases the soft 
excess is not accounted for, meaning that the soft excess may 
originate from another physical process.

\subsection{Investigation of relativistic reflection modelling over the
  0.3--10\,keV energy range}\label{sec:allpnref} 
 
We now consider the full 0.3--10\,keV energy range 
to investigate whether reflection models
 can after all reproduce both the soft excess and the hard X-ray emission up to
10\,keV. 

\begin{figure}[t!]
\begin{tabular}{c}
\includegraphics[width=1\columnwidth,angle=0]{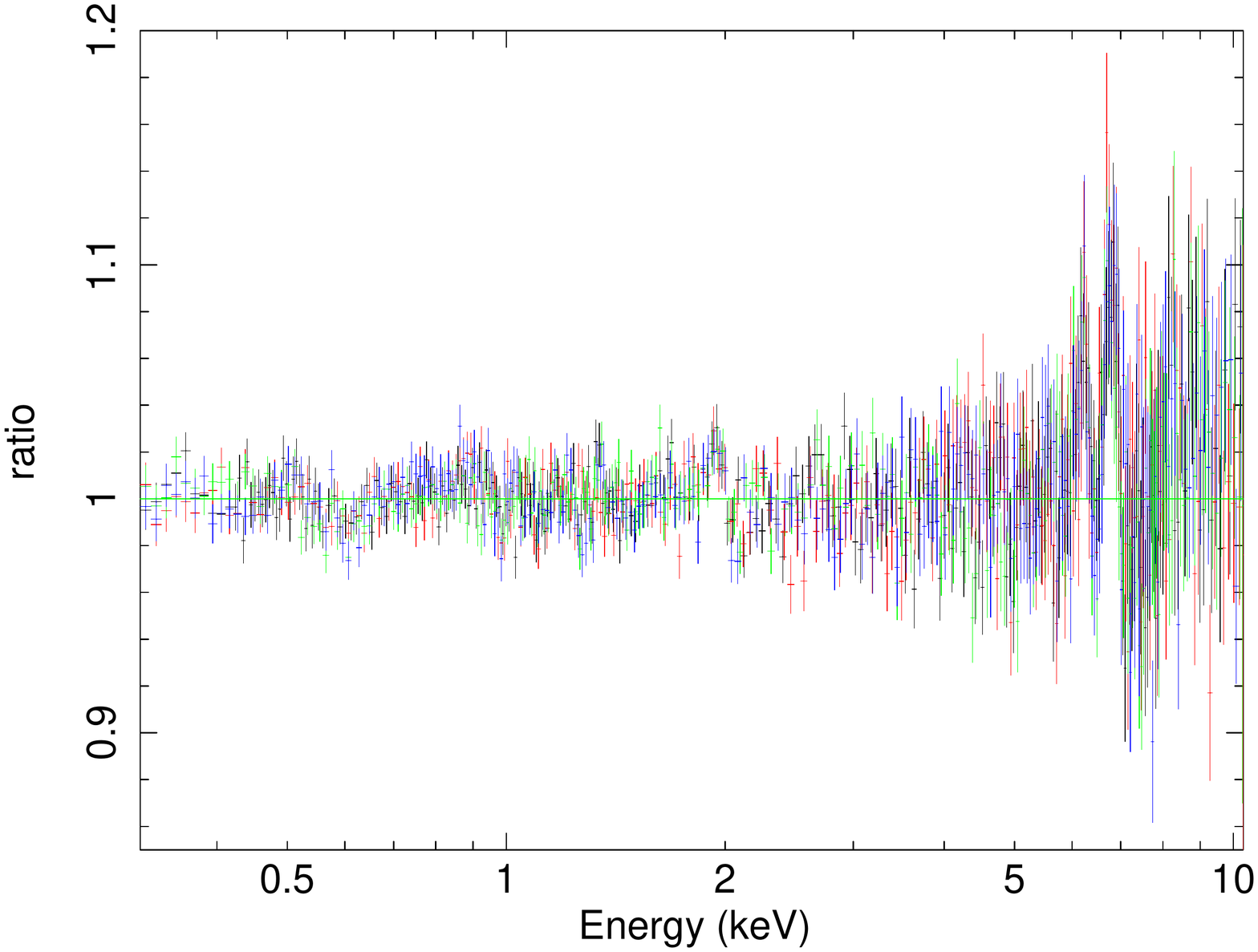}\\
\includegraphics[width=1\columnwidth,angle=0]{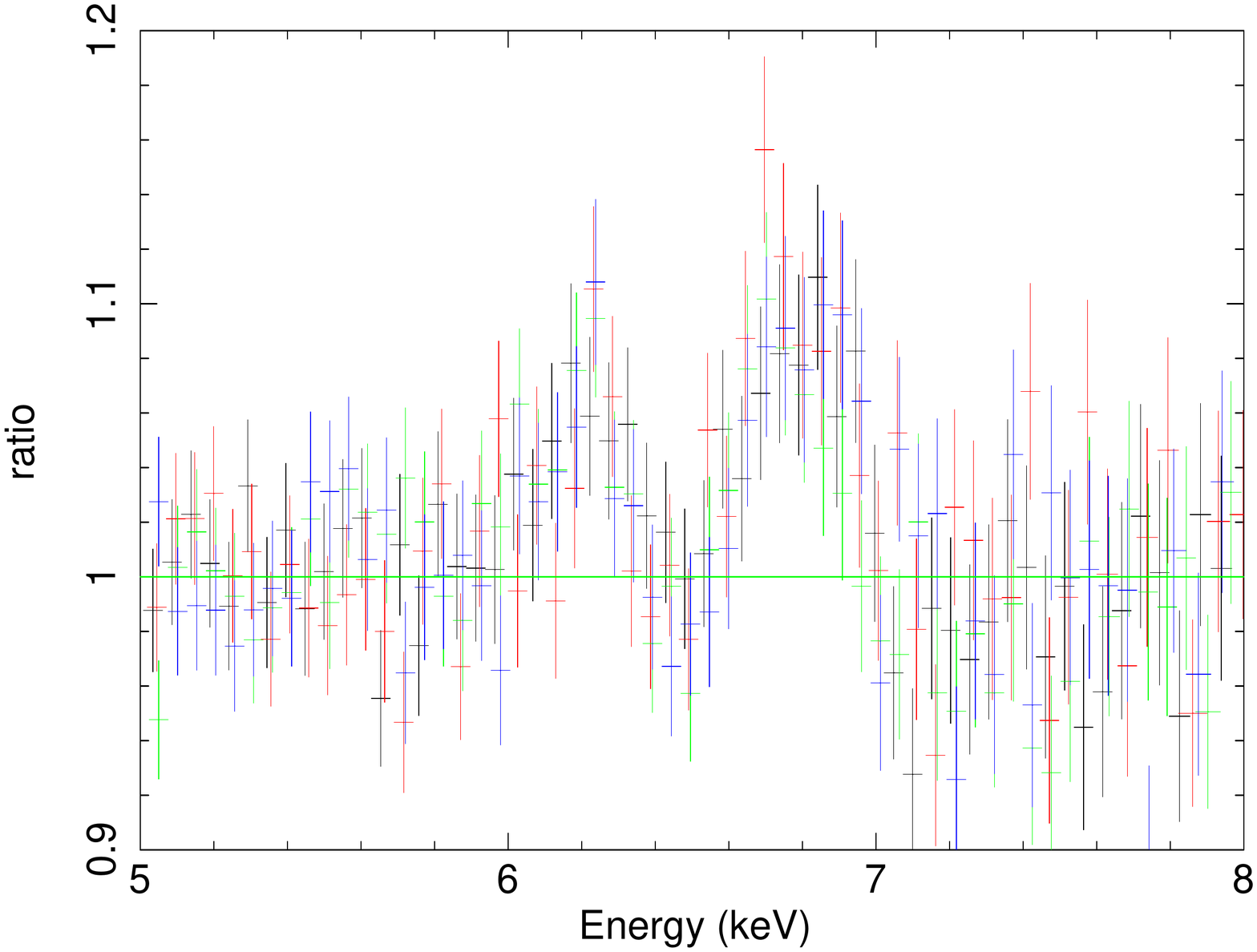}\\
\end{tabular}
\caption{Data/model ratio of the fit over the 0.3--10\,keV energy range 
of the four 2014 {\sl XMM-Newton}/pn spectra with the baseline relativistic
reflection model (model $\mathcal{A}$). 
The best fit parameters are reported in Table~\ref{tab:pnred}.  
Black: March 18, red: March 20,
 green: March 22, and blue: March 24. 
{\it Top panel}: The full 0.3--10\,keV energy range. 
{\it Bottom panel}: Zoom on the Fe\,K$\alpha$ complex.}  
\label{fig:relxill}
\end{figure}

We use model $\mathcal{A}$, but here we allow for a broken
power-law disk emissivity index ($q_1$, $q_2$, and $R_{\rm br}$). 
We find a good fit statistic ($\chi^{2}$/d.o.f.=7246.6/6728,
$\chi^{2}_{\rm red}$ = 1.08) over the 0.3--10\,keV energy range
(Fig.~\ref{fig:relxill}, top panel),
though there are positive residuals in the Fe\,K complex energy range
($\chi^{2}_{\rm red}$ = 1.16 when considering only the 6--7\,keV energy range). 
 Since the fit is driven by the smooth soft X-ray emission, very high values for
the spin and the disk emissivity indices, with a low inclination angle, are
required to reproduce it (see Table~\ref{tab:pnred}). 
Large values ($R$$\sim$7--10) for the reflection fraction are
required as well. This would correspond to a scenario of a compact corona
very close to the black hole, leading to strong gravitational
light bending \citep[][but see \citealt{Dovciak15}]{Fabian12}. 
Moreover the primary photon index needed to reproduce the soft X-ray 
excess is much steeper ($\Gamma$$\sim$2.4) than that associated to the Fe\,K$\alpha$
features when fitting above 3\,keV ($\Gamma$$\sim$1.9). 
These extreme parameters are incompatible with those found to reproduce the
3--10\,keV spectra (see Table~\ref{tab:pn3-10}). \\

Moreover, as shown in Figure~\ref{fig:relxill} (top panel), there are still 
positive residuals with moderately broad line profiles from about
6--6.3keV (red wing) and 6.6--6.9\,keV (blue wing) 
in all the four spectra (AGN rest-frame), which cannot be reproduced by the extreme and very
fine-tuned parameter values required to reproduce the smooth soft excess. 
The two emission features correspond to the red and blue Fe\,K$\alpha$ features
reported in Paper\,II, where the energy--time map showed that they both arise 
from the accretion disk. 
We therefore add two relativistic line components ({\sc relline},
\citealt{Dauser13}) that enable us to model these features for a small
inclination angle\footnote{We notice 
  that we find similar results considering a single
  relativistic line, but in such case a larger inclination angle of
  about 30 degrees is required (see Paper\,II, and
  Table~\ref{tab:pn3-10}).} 
 as found for the soft excess (see Table~\ref{tab:pn3-10}). 
 The inclination angle and the spin values are linked to
those of the broadband blurred reflection component, but we allow the
disk emissivity
index to vary ($R_{\rm in}$ is fixed to the ISCO).  
We find a statistical improvement of
the fit compared to the baseline model $\mathcal{A}$ 
($\Delta$$\chi^{2}$$\sim$$-$238  for five additional parameters). 
We infer line energies of 6.47$\pm$0.01\,keV
and 6.76$^{+0.02}_{-0.03}$\,keV, and a disk emissivity index 
 of 1.6$^{+0.2}_{-0.1}$, much flatter than the ones found for the
baseline reflection model $\mathcal{A}$, i.e., 7.5$\pm$0.5 and 4.5$\pm$0.4 for
the 2014 March 18 observation (similar values are found for the three other
sequences). If, instead, we force the disk emissivity index to be the same of the 
blurred component and allow the inner radius to vary, we also find a good fit 
($\Delta$$\chi^{2}$$\sim$$-$240 for five additional parameters)
with $R_{\rm in}$=25.7$^{+5.0}_{-4.3}$\,$R_{\rm g}$ and line 
energies\footnote{(p) means that the value is 
  pegged at the maximum or minimum ones allowed by the model.} of  
6.45$^{+0.10}_{-0.05(p)}$\,keV and 6.97$^{+0.00(p)}_{-0.05}$\,keV. 
Hence, in both cases, a flat disk emissivity index ($q$$\sim$1.6) or a large
inner radius ($R_{\rm in}$$\sim$25\,$R_{\rm g}$) are 
in conflict with the values required to account for the soft excess.  
This implies that this reflection model cannot self-consistently
produce both the soft excess and the mildly  
relativistic Fe\,K$\alpha$ line(s).
We also notice that, in order to reproduce these Fe\,K$\alpha$
 features by means of another broadband reflection component ({\sc relxill}), 
 not only $q$ (or $R_{\rm
  in}$) and the disk inclination must be untied, but also the $\Gamma$ value, which 
  must be much harder, i.e. below about 1.9.

So it is worth investigating whether these accretion disk features 
(soft excess and Fe\,K$\alpha$ residuals) can originate from a more
complex disk emissivity 
shape such as a twice broken power-law, which allows for an intermediate region 
with a flat emissivity index as could be found for a 
co-rotating continuum point source or an extended corona (see Figure 7 in
\citealt{wilkins12}). However, even with this reflection model we
cannot simultaneously reproduce these emission features.  
A disk ionization gradient (log\,$\xi$\,$\propto$\,$R^{-index}$)
  does not work either.
The inferred parameters for these fits are reported in Table~\ref{tab:pn}.  
Then, we test a lamppost geometry, but a less good fit is found
compared to the extended corona model ({\sc  relxill}), and
significant residuals near 6.35\,keV and 6.7\,keV are still present
and cannot be removed by allowing for 
an ionization gradient (assuming an $\alpha$ accretion disk or
 a powerlaw density profile
(density\,$\propto$\,$R^{\rm -index}$)). 
See Table~\ref{tab:pn} for the inferred fit parameters.

In conclusion, these high S/N spectra of Ark\,120 show that relativistic 
  reflection models from a constant-density, geometrically flat accretion disk, 
  while leading to reasonable $\chi^2/d.o.f$ value over 0.3--10\,keV, are 
  physically unsatisfactory, since they cannot simultaneously account for both the soft
  excess and the Fe\,K$\alpha$ lines. Note that such residual Fe\,K features could have
  been missed or readily neglected in lower S/N spectra, thus leading 
  to an interpretation of a relativistic reflection-dominated spectrum,
  with rather extreme and tuned parameters in terms of spin and emissivity.

\begin{table}[t!]
\caption{Simultaneous fit of the four 2014 {\sl XMM-Newton}/pn spectra with the
    baseline relativistic reflection 
	model (model $\mathcal{A}$) over the 0.3--10\,keV energy range. }             
\centering                          
\begin{tabular}{lcc}        
\hline\hline                 
Parameters & \multicolumn{1}{c}{relxill} & \multicolumn{1}{c}{relxilllp} \\    
\hline                       
$N_{\rm H}$ ($\times$10$^{21}$\,cm$^{-2}$) &  1.09$\pm$0.02  & 1.00$\pm$0.01 \\      
$a$         & 0.965$^{+0.003}_{-0.002}$ & 0.968$^{+0.003}_{-0.005}$  \\
$\theta$ (degrees)     &  $\leq$6.7 &$\leq$5.9\\
log\,$\xi$ (erg\,cm\,s$^{-1}$)       & $\leq$0.1 & $\leq$0.1\\
$A_{\rm Fe}$      &$\leq$0.5 & $\leq$0.5 \\
\hline                                  
                          &   \multicolumn{2}{c}{2014 March 18} \\
\hline                                  
$\Gamma$  & 2.48$\pm$0.02 & 2.37$\pm$0.01\\
$\mathcal{R}$       &  10.8$^{+1.1}_{-0.9}$ & $-$\\
$norm$  & 2.0$\pm0.1\times$10$^{-4}$ & 1.5$\pm0.1\times$10$^{-2}$ \\
$q_1$ &  7.4$^{+0.6}_{-0.5}$ & $-$\\
$q_2$ & 4.8$\pm$0.5  & $-$\\
$R_{\rm br}$ ($R_{\rm g}$) & 3.8$^{+1.0}_{-0.5}$ & $-$\\
$h$ ($R_{\rm g}$)   & $-$ & 1.8$\pm$0.1 \\
\hline                                  
                &   \multicolumn{2}{c}{2014 March 20} \\
\hline                                  
$\Gamma$      & 2.37$\pm$0.02 & 2.29$\pm$0.01\\
$\mathcal{R}$            & 7.5$\pm$0.7  & $-$\\
$norm$       & 1.7$\pm0.1\times$10$^{-4}$  & 1.0$\pm0.1\times$10$^{-2}$  \\
$q_1$          &6.8$^{+0.7}_{-0.6}$     & $-$ \\
$q_2$         & 3.4$^{+0.4}_{-0.3}$  & $-$\\
$R_{\rm br}$ ($R_{\rm g}$) & 5.2$^{+1.3}_{-1.0}$ & $-$\\
$h$ ($R_{\rm g}$)    & $-$ & 1.9$\pm$0.1  \\
\hline                                 
                &   \multicolumn{2}{c}{2014 March 22} \\
\hline                                  
$\Gamma$  &  2.38$\pm$0.02 & 2.27$\pm$0.01\\
$\mathcal{R}$    & 8.3$^{+0.9}_{-0.8}$    &$-$\\
$norm$    & 1.9$\pm0.1\times$10$^{-4}$ & 1.0$\pm0.1\times$10$^{-2}$  \\
$q_1$    &7.6$^{+0.8}_{-0.5}$   & $-$ \\
$q_2$    & 5.0$\pm$0.5  & $-$\\
$R_{\rm br}$ ($R_{\rm g}$) & 3.4$^{+0.8}_{-0.5}$ & $-$\\
$h$ ($R_{\rm g}$)   & $-$ & 1.9$\pm$0.1    \\
 \hline
                &   \multicolumn{2}{c}{2014 March 24} \\
\hline                                  
$\Gamma$ & 2.37$\pm$0.02  & 2.27$\pm$0.01  \\
$\mathcal{R}$    & 8.2$^{+0.8}_{-0.7}$   & $-$\\
$norm$ & 1.7$\pm0.1\times$10$^{-4}$ & 1.0$\pm0.1\times$10$^{-2}$  \\
$q_1$ &  8.1$^{+0.7}_{-1.0}$   & $-$ \\
$q_2$  & 4.3$^{+0.5}_{-0.6}$  & $-$\\
$R_{\rm br}$ ($R_{\rm g}$) & 3.6$^{+1.7}_{-0.5}$ & $-$\\
$h$ ($R_{\rm g}$)   & $-$ & 1.8$^{+0.1}_{-0.3}$ \\
 \hline
 \hline
$\chi^{2}$/d.o.f.  &  7246.6/6728 & 7507.4/6740 \\
$\chi^{2}_{\rm red}$ & 1.08  &  1.11 \\
\hline    \hline                  
\end{tabular}
\label{tab:pnred}
\end{table}

\subsection{Comptonization process as the origin of the soft excess}\label{sec:comptt4PN}

\begin{table}[t!]
\caption{Simultaneous fit of the four 2014 {\sl XMM-Newton}/pn data 
with the  baseline Comptonization model with {\sc Comptt} (model
$\mathcal{B}$, column 2)  over the
0.3--10\,keV energy range. The fit results when adding a {\sc
  relline} component are reported in column 3. {\sl nc} means that the
parameter value is not constrained.}             
\label{table:1}    
\centering                          
\begin{tabular}{@{}l c c }       
\hline\hline                 
Parameters & model$\mathcal\,{B}$ & + 1 relline\\  
\hline                       
$N_{\rm H}$ ($\times$10$^{21}$\,cm$^{-2}$)   &  1.06$\pm$0.01 & 1.07$\pm$0.01 \\   
$q$ & $-$ &  3(f) \\ 
$R_{\rm in}$ ($R_{\rm g}$) & $-$ &  45.3$^{+13.1}_{-11.0}$  \\ 
$a$ & $-$ & {\sl nc}   \\
$\theta$ (degrees) & $-$ & 30(f)  \\
$E$ (keV) &  $-$  & 6.52$^{+0.04}_{-0.03}$   \\
norm ($\times$10$^{-5}$) & $-$   & 2.7$\pm$0.3   \\
\hline                                  
                &   \multicolumn{2}{c}{2014 March 18}  \\
\hline                                  
$kT_{\rm e}$ (keV)& 0.84$^{+0.18}_{-0.17}$& 0.87$^{+0.30}_{-0.16}$\\
$\tau$ &  6.6$^{+0.8}_{-0.7}$ & 6.3$^{+0.6}_{-0.9}$ \\
$norm$ & 2.7$^{+0.7}_{-0.5}$ & 2.7$^{+0.6}_{-0.4}$ \\
$\Gamma$ & 1.78$^{+0.02}_{-0.03}$ & 1.79$\pm$0.03 \\
$norm$ ($\times$10$^{-2}$) & 1.1$\pm$0.1 & 1.1$\pm$0.1 \\
\hline                                  
                &   \multicolumn{2}{c}{2014 March 20} \\
\hline                                  
$kT_{\rm e}$ & 0.65$\pm$0.09 & 0.69$^{+0.08}_{-0.10}$ \\
$\tau$ & 8.0$^{+0.7}_{-0.5}$ & 7.7$^{+0.7}_{-0.6}$ \\
$norm$ & 2.1$\pm$0.3   & 2.0$\pm$0.3 \\
$\Gamma$ & 1.74$\pm$0.03 & 1.75$\pm$0.04 \\
$norm$ ($\times$10$^{-3}$)& 9.1$^{+0.6}_{-0.5}$    & 9.2$^{+0.6}_{-0.4}$  \\
\hline                                  
                &   \multicolumn{2}{c}{2014 March 22} \\
\hline                                  
$kT_{\rm e}$ & 0.79$\pm$0.14 & 0.83$^{+0.19}_{-0.14}$\\
$\tau$ & 7.1$^{+0.8}_{-0.6}$& 6.8$^{+1.1}_{-0.8}$ \\
$norm$ & 1.9$^{+0.4}_{-0.3}$ & 1.9$^{+0.4}_{-0.1}$ \\
$\Gamma$ & 1.74$^{+0.01}_{-0.03}$ & 1.75$^{+0.03}_{-0.04}$ \\
$norm$ ($\times$10$^{-2}$)  & 1.0$\pm$0.1 & 1.0$\pm$0.1 \\
 \hline
                &   \multicolumn{2}{c}{2014 March 24} \\
\hline                                  
$kT_{\rm e}$ & 0.63$^{+0.11}_{-0.09}$ & 0.68$^{+0.11}_{-0.12}$  \\
$\tau$ & 8.1$\pm$0.7 & 7.8$^{+0.8}_{-0.6}$ \\
$norm$ & 2.1$\pm$0.3 & 2.0$^{+0.4}_{-0.3}$ \\
$\Gamma$ & 1.74$\pm$0.03 & 1.74$\pm$0.04\\
$norm$ ($\times$10$^{-3}$) & 9.5$^{+0.5}_{-0.3}$ & 9.4$\pm$0.7 \\
 \hline
 \hline
$\chi^{2}$/d.o.f.& 7384.1/6735 & 7165.2/6732 \\
$\chi^{2}_{\rm red}$& 1.10 & 1.06 \\
\hline    \hline                  
\end{tabular}
\label{tab:4pncomptt}
\end{table}

\begin{figure}[t!]
\begin{tabular}{c}
\includegraphics[width=0.9\columnwidth,angle=0]{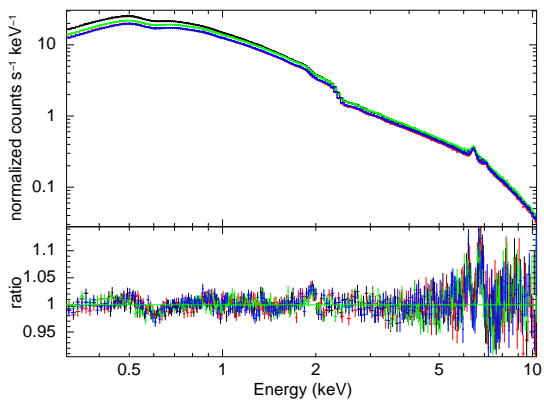}\\
\includegraphics[width=0.9\columnwidth,angle=0]{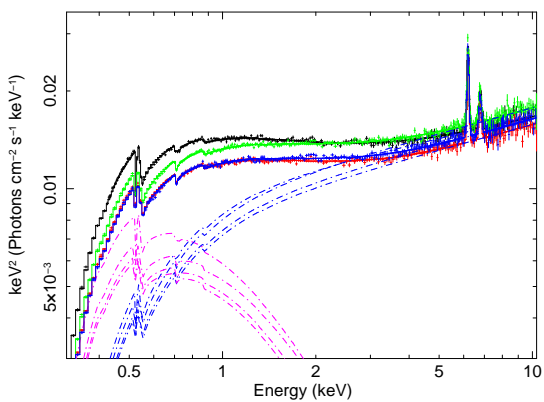}\\
\end{tabular}
\caption{Fit over the 0.3--10\,keV energy range 
of the four 2014 {\sl XMM-Newton}/pn spectra with the baseline Comptonization
model (model $\mathcal{B}$).  
The fit parameters are reported in Table~\ref{tab:4pncomptt}.  
Black: March 18, red: March 20,
 green: March 22, and blue: March 24. 
{\it Top panel:} data and data/model ratio. 
{\sl Bottom panel:} Unfolded spectra where the contribution of the model components are
displayed. The following colour code for the emission components
(dot-dashed curves) is used: magenta 
 for the soft Comptonization ({\sc comptt}), blue for the {\sc
   cut-off power law} (hot Comptonization). 
For clarity purposes, we have not displayed the Fe\,K line components.}  
\label{fig:pncompt}
\end{figure}

\begin{figure}[t!]
\begin{tabular}{c}
\includegraphics[width=0.9\columnwidth,angle=0]{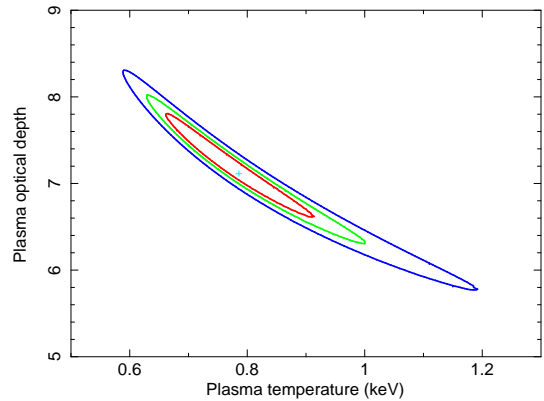}\\
\end{tabular}
\caption{2D contour plot  plot of the plasma
temperature ($kT_{\rm e}$ in keV) versus the plasma optical depth for
the third XMM-Newton observation inferred from the baseline Comptonization
model (model $\mathcal{B}$).}  
\label{fig:2Dcontour}
\end{figure}

In this section, we assume that the soft X-ray excess originates from
the Comptonization of seed photons from the accretion disk by
warm electrons from the corona (here using {\sc comptt}), as found by
\cite{Matt14} for the February 2013 
observation. For this, we use the baseline model $\mathcal{B}$ defined
as 
{\sc tbnew$\times$[comptt+zpo+3$\times$zgaussians(BLR)]}.  
 The {\sc comptt} model \citep{Titarchuk94} is characterized by
  the input soft photon temperature (expressed in keV), the plasma
  temperature ($kT_{\rm e}$ expressed in keV), the plasma optical depth
  ($\tau$) and the geometry assumed (disk, sphere, analytical 
  approximation). We assume a disk geometry, and 
an input soft photon temperature\footnote{We would like to notice
  that the mean value of the input soft photon temperature (around
  15\,eV) has a negligible impact on
  the plasma temperature and optical depth values.} of 15\,eV according to the black hole mass and
the mean accretion rate of Ark\,120. The power-law component 
 ({\sc zpo}) is used
to readily reproduce Comptonization by the hot electrons of the
corona. 
We obtain an overall good fit with $\chi^{2}$/d.o.f.=7384.1/6735 
($\chi^{2}_{\rm red}$=1.10) (see Table~\ref{tab:4pncomptt},
column 2), except for the
positive residuals at  the Fe\,K$\alpha$ complex energy range (see below and
Fig.~\ref{fig:pncompt} top panel), as already found in the previous reflection-based models. 
We derive for the Comptonized plasma low temperature values of about
0.6--0.8\,keV, and high optical depth values of about 7--8 
for the four observations (see the mean values and their associated
errors bars for each observation in Table~\ref{tab:4pncomptt}).
 For illustration purpose, we display in Fig.~\ref{fig:2Dcontour} the 2D contour plot of the plasma
temperature ($kT_{\rm e}$ in keV) versus the plasma optical depth for
the third XMM-Newton observation.
The temperature value of the four observations seems to follow the soft excess strength, but the
values between the four observations are consistent with each other
within their errors bars. The primary photon indices of about
 1.74--1.79 are much harder than those required to produce 
the soft excess from relativistic reflection modelling ($\Gamma$$\sim$2.4).

Not surprisingly with such a power-law continuum shape above 3\,keV,
the same red and blue components of Fe\,K$\alpha$ residuals are still present. 
Indeed, they are known to be associated with the accretion disk (Paper\,II). 
Thanks to the high S/N of the present spectra and the ability to detect 
unambiguously these red and blue emission features, 
we are thus able to establish that even in this scenario, 
where the soft excess originates from Comptonization, a relativistic
reflection component is still required. 
We therefore add a relativistic line profile ({\sc relline}), fixing its 
emissivity index to the standard value of 3 and the inclination angle to 
30\,degrees (see Paper\,II). 
 If $R_{\rm in}$ is allowed to vary, then the fit is significantly improved 
(see Table~\ref{tab:4pncomptt}, column 3), and $R_{\rm
  in}$=45.3$^{+13.1}_{-11.0}$\,$R_{\rm g}$ is inferred. 
The relative contribution of the different model components is displayed in
Fig.~\ref{fig:pncompt} (unfolded spectrum for illustration
purposes only; bottom panel), and shows 
that Comptonization of seed disk photons by warm electrons of the
corona is the dominant process below about 0.8\,keV. 

In summary, the spectral analysis shows that Comptonization by a
low temperature ($kT_{\rm e}$$\sim$0.6--0.8\,keV) optically-thick
($\tau$$\sim$7--8) corona can reproduce well the soft
excess below 1\,keV. Above this energy, the power law component
dominates the continuum, which may represent Comptonization by
the hot electrons in a thin corona \citep[e.g.,][]{Haardt93,Zdziarski95}. 
 We note that a contribution from disc reflection originating at tens
of $R_{\rm g}$ is still required to account for a broad component of the FeK
line, which was not the case for the low flux 2013 observation of
Ark 120 \citep{Matt14}.  

\section{Broad-band X-ray view of Ark 120 observed on 2014 March 22}\label{sec:broadbandX}

\begin{figure}[t!]
\begin{tabular}{cc}
\includegraphics[width=0.9\columnwidth,angle=0]{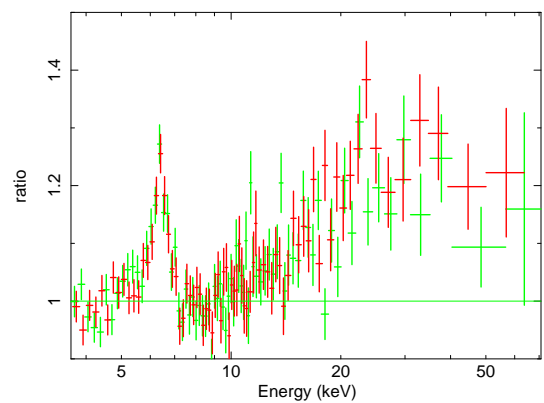} 
\end{tabular}
\caption{Data/model ratio of the {\sl NuSTAR} spectra of Ark\,120 obtained
  on 2014 March 22, fitted 
 with a Galactic absorbed power-law continuum model
 ($\Gamma$=1.92$\pm$0.02) over the
 3.5--10\,keV energy range (excluding the 5.5--7.5\,keV energy range,
 i.e. the Fe\,K complex) 
and then extrapolated over the 3.5--79\,keV
energy range. In addition to the prominent Fe\,K$\alpha$ complex, a
significant hard X-ray excess is present. 
Red: {\sl NuSTAR} FPMA, and green: {\sl NuSTAR} FPMB.}  
\label{fig:broad3-5keV}
\end{figure}

In this section, we investigate the simultaneous broadband {\sl
  XMM-Newton}/pn and 
{\sl NuSTAR} observations of Ark\,120 performed on 2014 March 22.
 
 First, we investigate the two {\sl NuSTAR} spectra (FPMA and
  FPMB) by fitting them 
 using a simple absorbed power-law model in the 3--10\,keV energy
 range excluding the 5.5--7.5\,keV energy range. We fix  
the absorption column density to 9.78$\times$10$^{20}$\,cm$^{-2}$, and
tie the power-law parameters between both {\sl NuSTAR} spectra. 
 We allow for possible cross-calibration uncertainties between these two {\sl NuSTAR}
spectra. The resulting data/model ratio extrapolated up to
  79\,keV is reported in Fig.~\ref{fig:broad3-5keV}, where a significant hard X-ray excess is present
in addition to the prominent Fe\,K$\alpha$ complex.

\subsection{Investigation of the relativistic reflection scenario}\label{sec:2014cref}

\subsubsection{Spectral analysis above 3\,keV}

\begin{table}[t!]
\caption{Simultaneous {\sl XMM-Newton}/pn and {\sl NuSTAR} spectral
  fit of the 2014 March 22 observation  
with the relativistic reflection model above 3\,keV (model $\mathcal{A}$). 
{\it nc} means that the parameter value is not constrained.  }        
\label{table:ref2014c}    
\centering                          
\begin{tabular}{l c c}       
\hline\hline                
Parameters &  \multicolumn{2}{c}{relxill} \\ 
\hline                       
  & \multicolumn{1}{c}{3--10\,keV}&    \multicolumn{1}{c}{3--79\,keV} \\
\hline
$q$  &2.0$^{+0.5}_{-0.2}$  & 2.2$^{+0.4}_{-0.2}$ \\
$a$ & {\it nc}  & $\leq$0.6 \\
$\theta$ (degrees) & $\leq$25.6  & 21.9$^{+2.8}_{-12.8}$ \\
$\Gamma$ & 1.91$^{+0.03}_{-0.02}$ & 1.86$^{+0.01}_{-0.02}$   \\
log\,$\xi$  (erg\,cm\,s$^{-1}$)& 2.8$^{+0.1}_{-0.2}$& 2.7$^{+0.1}_{-0.3}$ \\
$A_{\rm Fe}$ &  $\leq$0.8 & 3.3$^{+1.5}_{-1.3}$ \\
$E_{\rm cut}$(keV) & 1000 (f)  & 364$^{+320}_{-170}$  \\
$\mathcal{R}$  & 0.6$^{+0.1}_{-0.4}$  & 0.3$\pm$0.1  \\ 
norm ($\times$10$^{-4}$) & 2.1$^{+0.3}_{-0.2}$  & 2.4$\pm$0.3  \\
C$_{\rm NuSTAR\,A}$ & 1.030$\pm$0.009  &1.029$\pm$0.009 \\
C$_{\rm NuSTAR\,B}$ & 1.070$\pm$0.010  &1.072$\pm$0.009 \\
 \hline
 \hline
$\chi^{2}$/d.o.f. &1207.6/1232  & 1497.7/1518 \\
$\Delta\chi^{2}_{\rm red}$& 0.98 & 0.99 \\
\hline    \hline                  
\end{tabular}
\label{tab:2014cabove3keV}
\end{table}

\begin{figure}[t!]
\begin{tabular}{cc}
\includegraphics[width=0.9\columnwidth,angle=0]{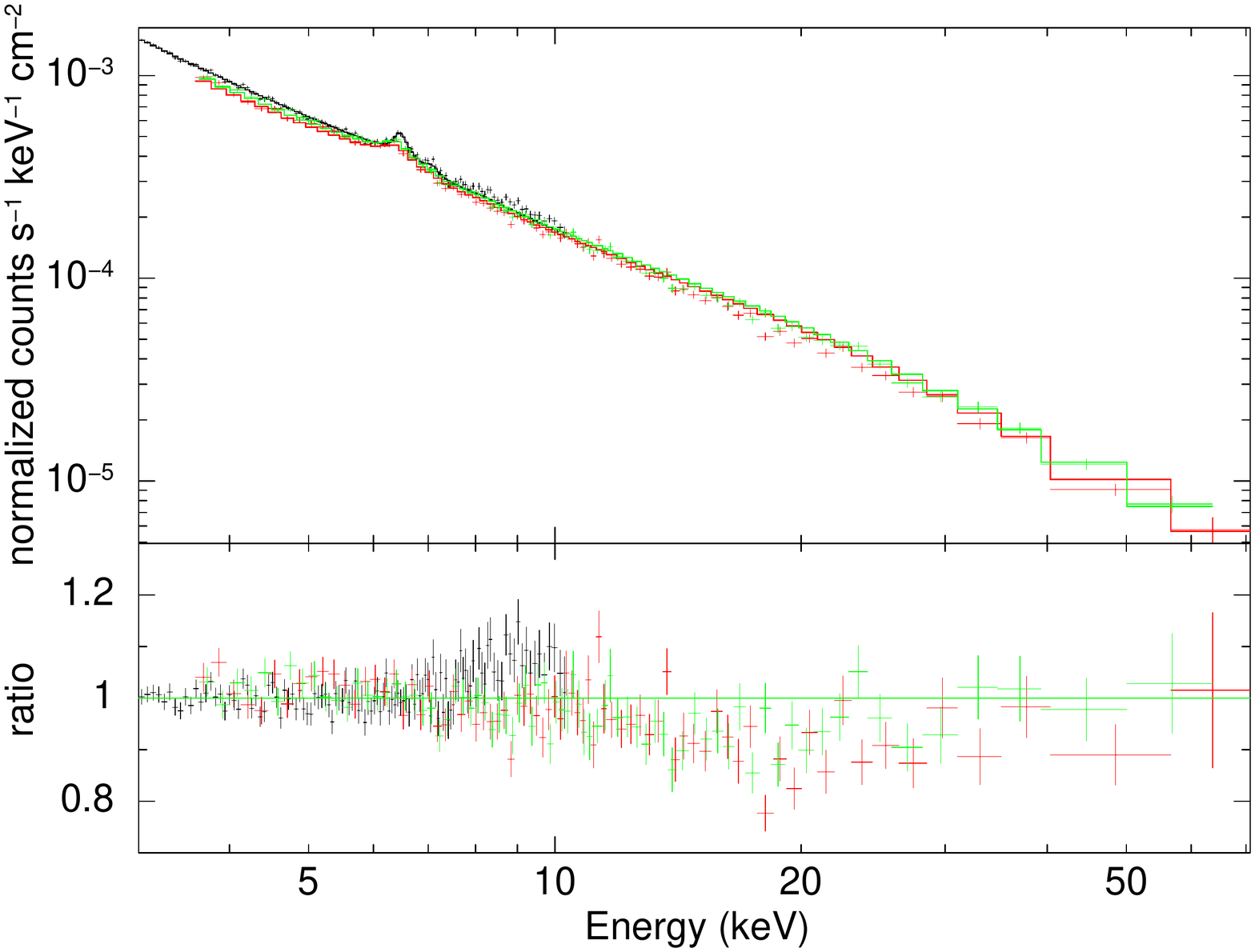}
\\
\includegraphics[width=0.9\columnwidth,angle=0]{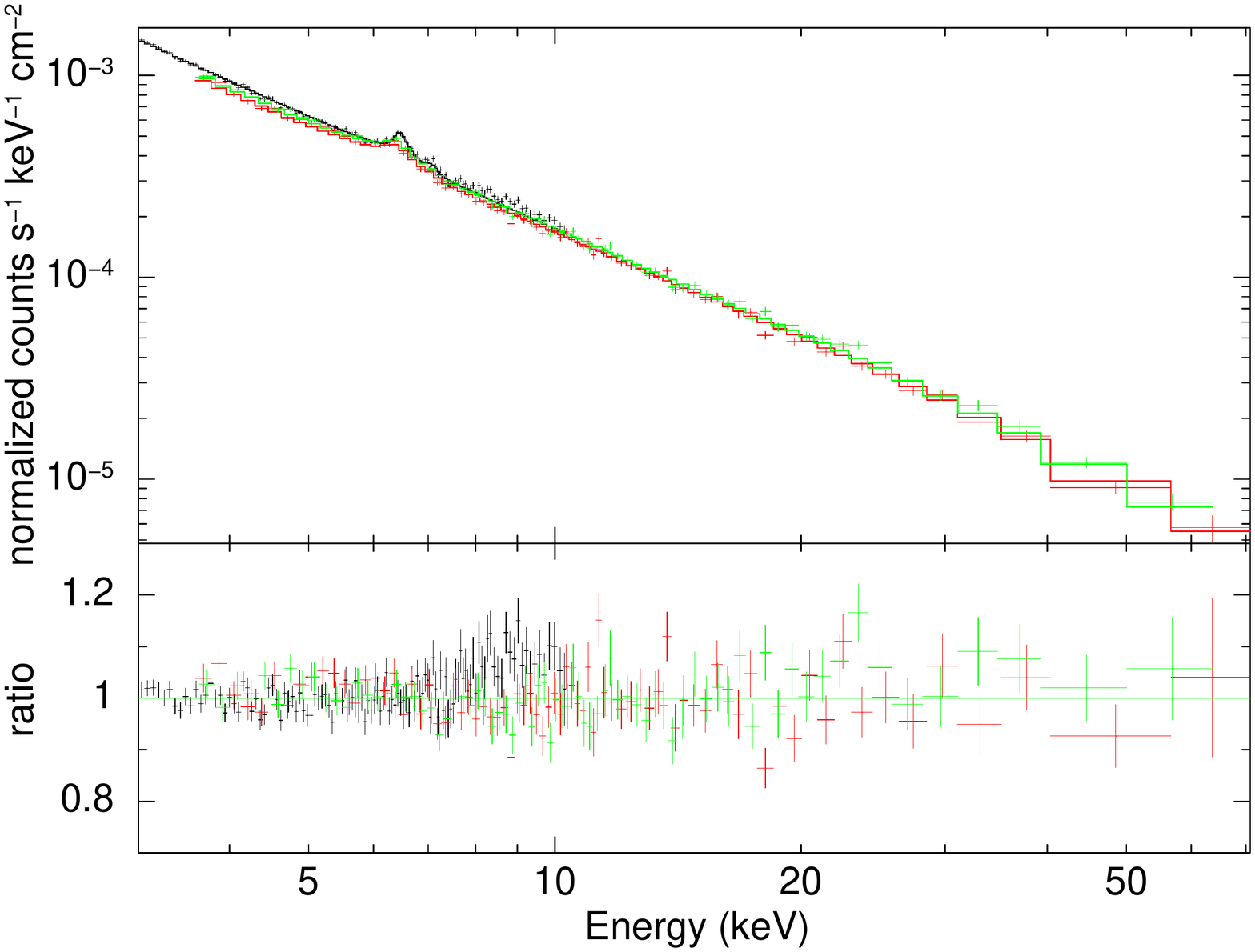}\\
\end{tabular}
\caption{The simultaneous {\sl XMM-Newton}/pn and {\sl NuSTAR} spectra
  of Ark\,120 obtained on 2014 March 22 fitted 
 with model $\mathcal{A}$.  
Black: {\sl XMM-Newton}/pn spectrum, red: {\sl NuSTAR} FPMA, and green:
{\sl NuSTAR} FPMB.  
{\it Top panel:} fit over the 3--10\,keV range and
 then extrapolated up to 79\,keV without refitting.
{\it Bottom panel:} fit over the 3--79\,keV range.}   
\label{fig:above3keV}
\end{figure}

We start the simultaneous fit of the {\sl XMM-Newton}/pn and
 of the two {\sl NuSTAR} spectra in the 3--10\,keV band 
using model\,$\mathcal{A}$.  
We allow for cross-calibration uncertainties between the two {\sl NuSTAR}
spectra and the {\sl XMM-Newton}/pn spectrum by including in
the fit a cross-normalization constant corresponding to 
C$_{\rm NuSTAR\,A}$ and C$_{\rm NuSTAR\,B}$ for NuSTAR FPMA and FPMB
spectra, respectively (see values in Table~\ref{tab:2014cabove3keV}), related to
the {\sl XMM-Newton}/pn spectrum. 
 The absorption column density has been fixed
to the Galactic one, i.e. 9.78$\times$10$^{20}$\,cm$^{-2}$. 
The fit parameters are similar (see
Table~\ref{tab:2014cabove3keV}, column 2) to those found when fitting
simultaneously the four {\sl XMM-Newton}/pn spectra over this energy range (see
Table~\ref{tab:pn3-10}), showing a good match between
{\sl XMM-Newton}/pn and {\sl NuSTAR} data. Only the inferred disk emissivity
index $q$ is larger and the inclination angle is lower 
due to the lower spectral resolution of the {\sl NuSTAR} 
camera which broadens the apparent Fe\,K profile. 
We then extrapolate this fit up to 79\,keV, and find that the
data/model ratio is rather good, although the model slightly
overpredicts the emission in the 10--40\,keV energy range (see
Fig.~\ref{fig:above3keV}, top panel).  
The $\chi^{2}$/d.o.f. is 1698.1/1519 ($\Delta\chi^{2}_{\rm
  red}$=1.12) without any refitting. Now, we refit over the entire 
3--79\,keV energy range and find a very good fit (Fig.~\ref{fig:above3keV}, bottom panel) 
with parameter values that are 
very similar to those found for the 3--10\,keV energy range,
except for the iron abundance which has increased from $\leq$0.8 to
3.3$^{+1.5}_{-1.3}$ in order to better adjust the 10--40\,keV
emission (see Table~\ref{tab:2014cabove3keV}).  
We find an upper limit of 0.6 for the spin value (at 90\% confidence
level), but we note that after the calculation of a 2D contour plot of the spin versus the
inclination angle, the spin is actually unconstrained at the 90\% confidence level. 
A power-law photon index of about 1.86 
represents well the underlying continuum over the 3--79\,keV energy
range. This value is softer than that found for the 2013 observation
($\Gamma$$\sim$1.73), when the source was in a low-flux state
\citep{Matt14}.  
Adding the possible contribution from a molecular torus\footnote{For
 this, we use the unblurred reflection {\sc xillver} model fixing
 log\,$\xi$ to zero. The inclination of the blurred and unblurred
 reflection components, as well as their cut-off energies, are tied together,
 and solar iron abundance for the unblurred reflection
 component is assumed. }  
 does not improve the fit at all. 

\begin{figure}[t!]
\begin{tabular}{cc}
\includegraphics[width=0.9\columnwidth,angle=0]{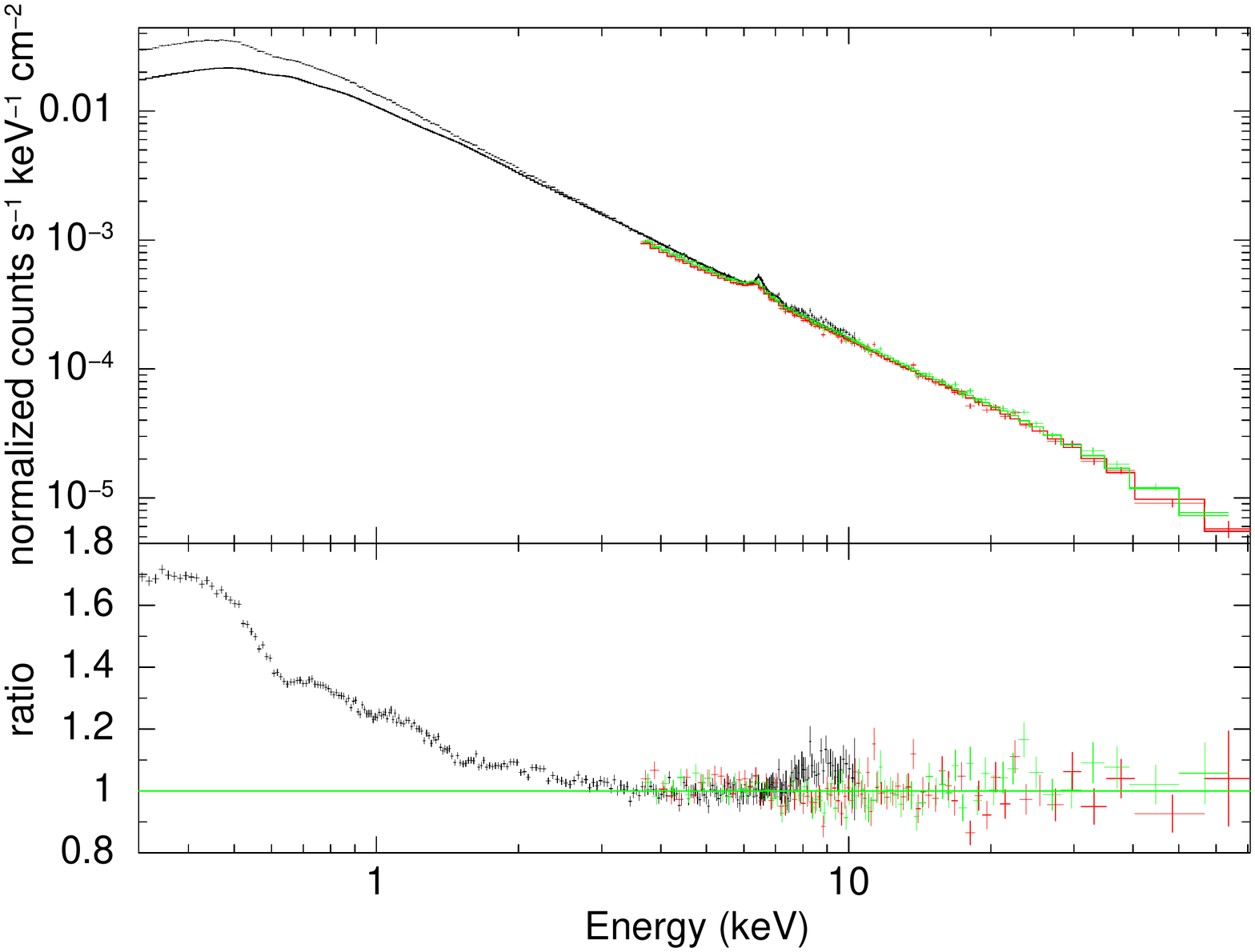}
\\
\includegraphics[width=0.9\columnwidth,angle=0]{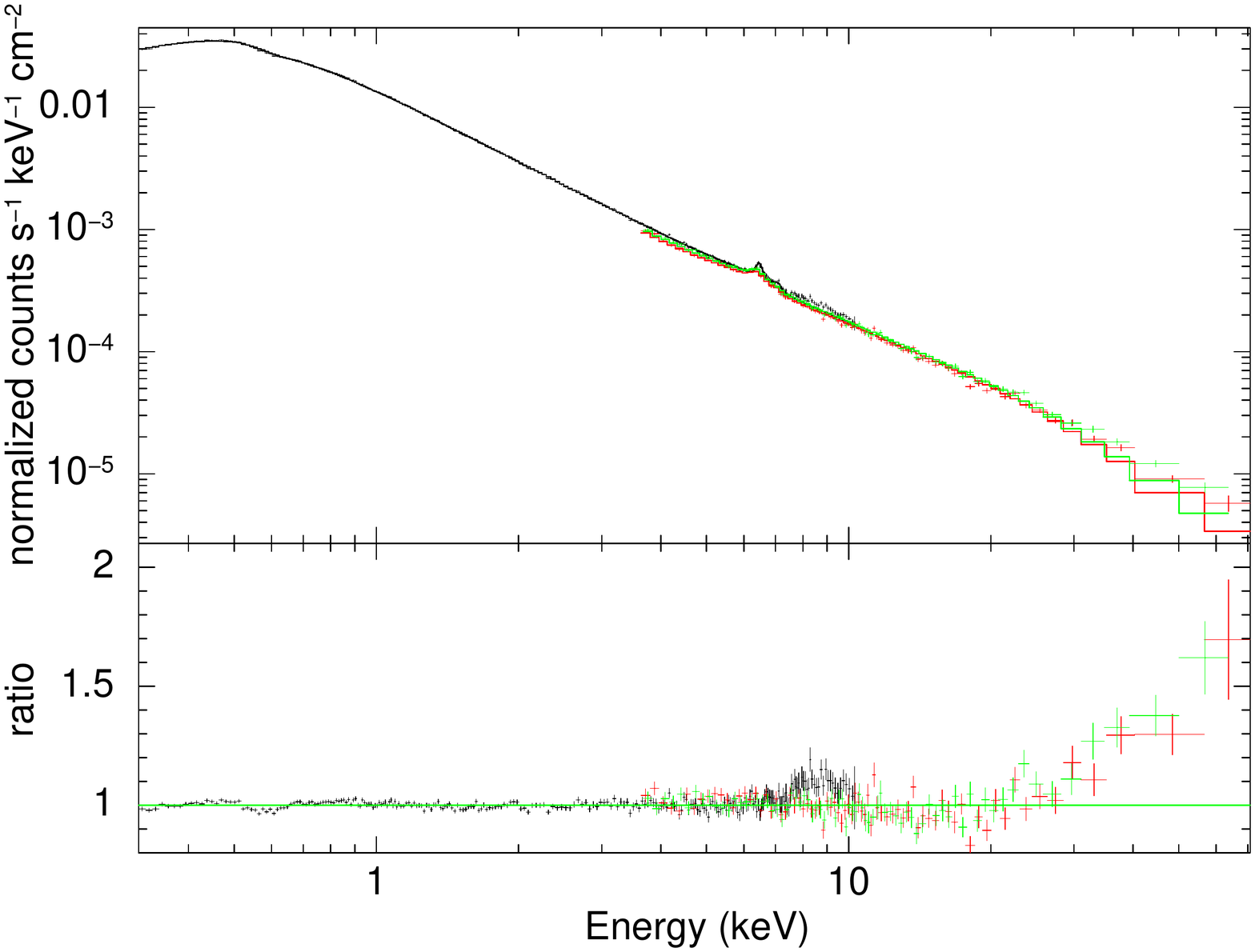} 
\end{tabular}
\caption{The simultaneous {\sl XMM-Newton}/pn and {\sl NuSTAR} spectra
  of Ark\,120 obtained  on 2014 March 22 fitted 
 with model $\mathcal{A}$ assuming a coronal geometry. {\it Top panel}: Fit over 
 the 3--79\,keV energy range (see
 Table~\ref{tab:2014cabove3keV}). The extrapolation down to 0.3\,keV 
 clearly shows that the soft excess is not accounted for.
{\it Bottom panel}: Fit over the 0.3-79\,keV energy range.  The corresponding
fit parameters are reported in Table~\ref{tab:2014cref}. 
The reflection model clearly leaves a strong hard X-ray excess above
30\,keV and
thus cannot account simultaneously for the soft and hard bands. 
Black: {\sl XMM-Newton}/pn spectrum, red: {\sl NuSTAR} FPMA, and green:
{\sl NuSTAR} FPMB.}   
\label{fig:extrapolation}
\end{figure}

As illustrated in Fig.~\ref{fig:extrapolation} (top panel), the
extrapolation of the {\sl XMM-Newton}/pn spectrum down
to 0.3\,keV shows that this model is not able to account for the
soft X-ray excess, as might be expected given that the best fit 
is characterized by similar parameters as when considering the 
3--10\,keV band alone. \\

\subsubsection{Spectral analysis over the 0.3--79\,keV energy range}\label{sec:nustarall}

First, we note that, if we try to reproduce the hard X-ray excess with
a contribution from the torus using {\sc xillver},\footnote{Same results
are found if, instead of {\sc  xillver},  
we use {\sc mytorus} \citep{Murphy09} or {\sc pexmon}
\citep{Nandra07}.}  
then the normalization of the Fe\,K$\alpha$ line emitted by the BLR is
consistent with 0. 
This, is at odds with analysis of the
{\sl Chandra}/HETG spectra, where the narrow profile is resolved with a width
compatible with the BLR (Paper\,II). We can therefore conclude, from both the 
3--79\,keV and 0.3--79\,keV analysis, that any contribution from the
torus is not significant, and is not considered from now on.
However, we checked that the following results do not depend on the modeling of the
narrow Fe\,K$\alpha$ core either from the BLR (as established) in paper\,II or
by the common modeling of a molecular torus. 

\begin{table}[t!]
\caption{Simultaneous {\sl XMM-Newton}/pn and {\sl NuSTAR} spectral
  fit of the 2014 March 22 observation with the relativistic reflection model (model $\mathcal{A}$) 
over the broad 0.3--79\,keV energy range.  
{\it nc} means that the parameter value is not constrained. 
 }            
\label{table:ref2014c}    
\centering                         
\begin{tabular}{l  c c }       
\hline\hline                
      &  \multicolumn{2}{c}{0.3--79\,keV}  \\
Parameters & \multicolumn{1}{c}{relxill} & \multicolumn{1}{c}{relxilllp}\\    
\hline                       
$N_{\rm H}$ ($\times$10$^{21}$\,cm$^{-2}$)&  0.93$\pm$0.01 & 0.93$\pm$0.02   \\      
$q_1$  & $\geq$8.3  & $-$\\
$q_2$ & 3.0$\pm$0.2 & $-$ \\
$R_{\rm br}$ ($R_{\rm g}$) & 2.6$\pm$0.2 & $-$\\
$h$ ($R_{\rm g}$) & $-$ &  2.2$\pm$0.1  \\
$a$ &      0.989$^{+0.003}_{-0.006}$ & $\geq$0.993 \\
$\theta$ (degrees) & 34.2$^{+1.7}_{-2.3}$  & 35.0$^{+1.9}_{-2.2}$ \\
$\Gamma$ &2.22$\pm$0.01 & 2.21$\pm$0.01 \\
log\,$\xi$  (erg\,cm\,s$^{-1}$) & $\leq$0.1   & $\leq$0.1  \\
$A_{\rm Fe}$ &  1.2$\pm$0.3 & 1.2$\pm$0.2 \\
$E_{\rm cut}$(keV) & $\geq$956 &  $\geq$801\\
$\mathcal{R}$  & 4.1$^{+0.3}_{-0.6}$ &$-$ \\ 
norm & 1.9$\pm$0.1$\times$10$^{-4}$ & 3.8$^{+0.7}_{-0.5}$$\times$10$^{-3}$ \\
C$_{\rm NuSTAR\,A}$ & 1.031$\pm$0.008 & 1.026$\pm$0.009\\
C$_{\rm NuSTAR\,B}$ & 1.075$\pm$0.009 & 1.070$\pm$0.009\\
 \hline
 \hline
$\chi^{2}$/d.o.f. &  2483.2/2058 & 2585.3/2061 \\
$\chi^{2}_{\rm red}$& 1.21  & 1.25\\
\hline    \hline                  
\end{tabular}
\label{tab:2014cref}
\end{table}

Using model $\mathcal{A}$ (and allowing for a single broken power-law
emissivity index), 
we are able to find a satisfactory fit only up to 30\,keV, with the parameters reported in 
Table~\ref{tab:2014cref} ($\chi^{2}$/d.o.f.=2483.2/2058,
$\chi^{2}_{\rm red}$=1.21). Indeed, as shown in
Figure~\ref{fig:extrapolation} (bottom panel), above
 this energy there is a significant hard X-ray excess.
Of course, as for the four {\sl XMM-Newton}/pn spectra, there are
still residuals present at Fe\,K$\alpha$, but these appear less apparent on the
model/ratio plot due to the very significant positive residual
observed above about 30\,keV. 
As also found for the 0.3--10\,keV spectral
analysis, a large reflection fraction and a very steep disk emissivity shape ($q_1$$\gtrsim$8 or
$h$$\sim$2\,$R_{g}$) are required, 
as well as a very high (and strongly fine-tuned) value for the spin.  
 Moreover, the inferred photon index
is significantly steeper ($\Gamma$$\sim$2.2) than that found
considering only data above 3\,keV (i.e., $\Gamma$$\sim$1.9) 
 explaining the presence of the hard X-ray excess residual seen above about
 30\,keV. 
We also notice that this hard X-ray excess residual cannot be accounted for
by any other alternative {\sc relxill} models (i.e., the ones tested
in $\S$\ref{sec:allpnref}). 
The lamppost geometrical configuration leads to the 
worst fit of the data (see Table~\ref{tab:2014cref}) 
We check whether such unsatisfactory fits are due to the specific model 
components and assumptions, still using the baseline model 
$\mathcal{A}$ by alternatively: 
\begin{itemize}
\item Allowing the primary photon index of the {\sl NuSTAR} spectra to be 
  different from that of the {\sl XMM-Newton}/pn spectrum in order to compensate for 
  any possible calibration issues.   
\item  Replacing the {\sc relxill} model with {\sc kyreflionx}, which is a model combining the
    relativistic smearing \citep{Dovciak04} 
and X-ray ionized reflection models {\sc reflionx} \citep{Ross05} or
{\sc xillver} \citep{Garcia13}. 
\item Allowing the incident continuum of the relativistic 
  component to be different from the direct, observed one (see
  appendix~\ref{sec:complex} for details)
\item Assuming a larger accretion disk density of 10$^{19}$\,cm$^{-3}$ 
(see appendix~\ref{sec:highdens} for details)
\end{itemize}
Yet, none of these reflected dominated scenarios allows us to account
  simultaneously for the 3 main components (soft excess, broad
 Fe\,K$\alpha$ lines and Compton hump), 
 or to obtain physically meaningful fit results. 
Furthermore, limiting the analysis to one or another energy range
would lead to erroneous results on the physical condition of the
disk/corona system (see a comparison of Table~\ref{tab:2014cabove3keV}
and Table~\ref{tab:2014cref}).  \\

To summarize, during this 2014 observational campaign of Ark\,120, we can
safely rule out relativistic reflection as the origin of both the soft and hard 
X-ray excesses, and the red/blue relativistic Fe\,K$\alpha$ features.

\subsection{Model combining Comptonization and relativistic reflection}\label{sec:2014ccomp}

Here we investigate if a combination of soft and hard Comptonization 
and mildly relativistic reflection can explain the whole
0.3--79\,keV continuum shape, as found for the four {\sl
  XMM-Newton}/pn spectra. To do this, we use model $\mathcal{C}$
defined as 
{\sc tbnew$\times$[comptt+cut-off PL+relxill+3$\times$zgaussians(BLR)]}. 

\begin{figure}[t!]
\begin{tabular}{c}
\includegraphics[width=0.9\columnwidth,angle=0]{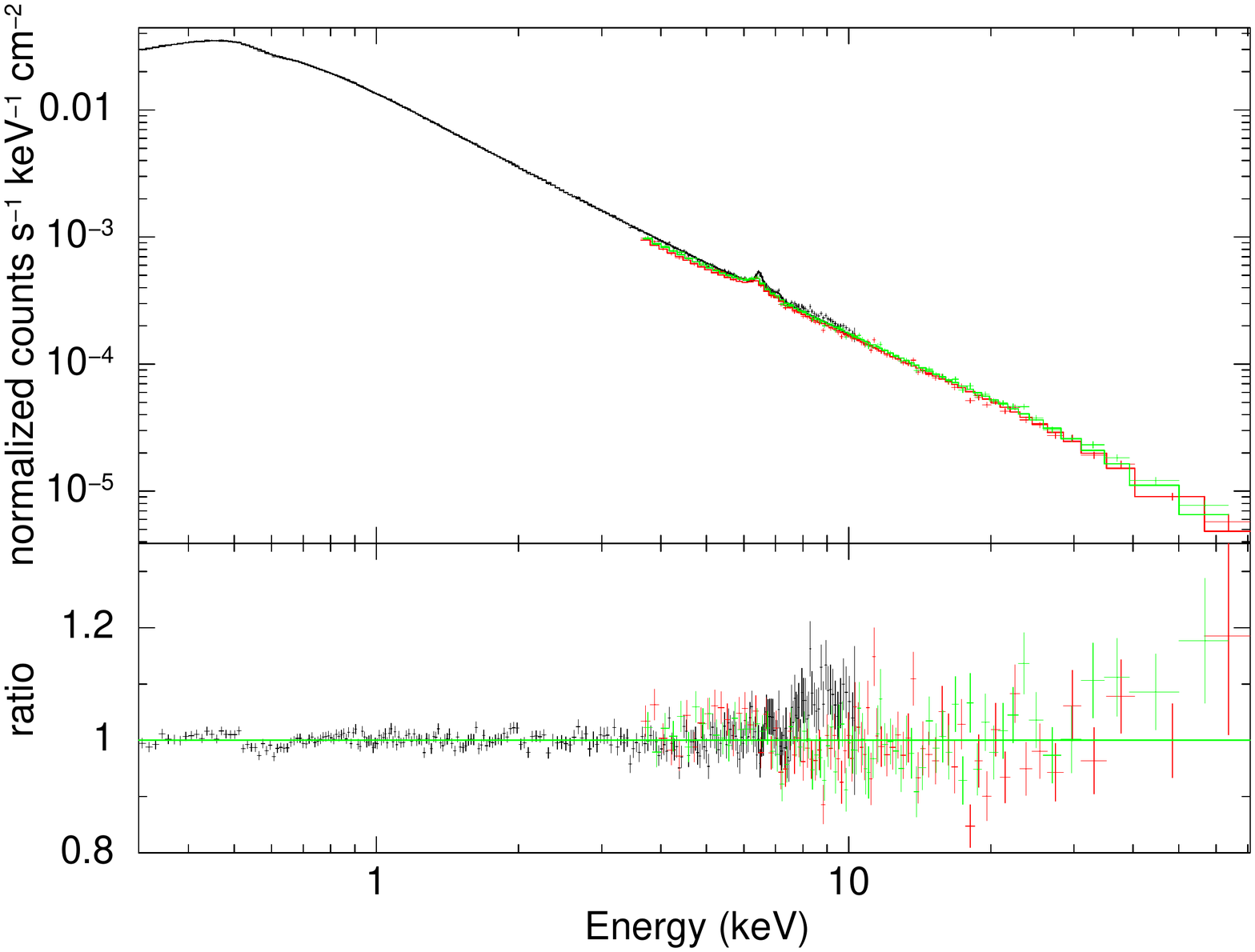}  \\
\includegraphics[width=0.9\columnwidth,angle=0]{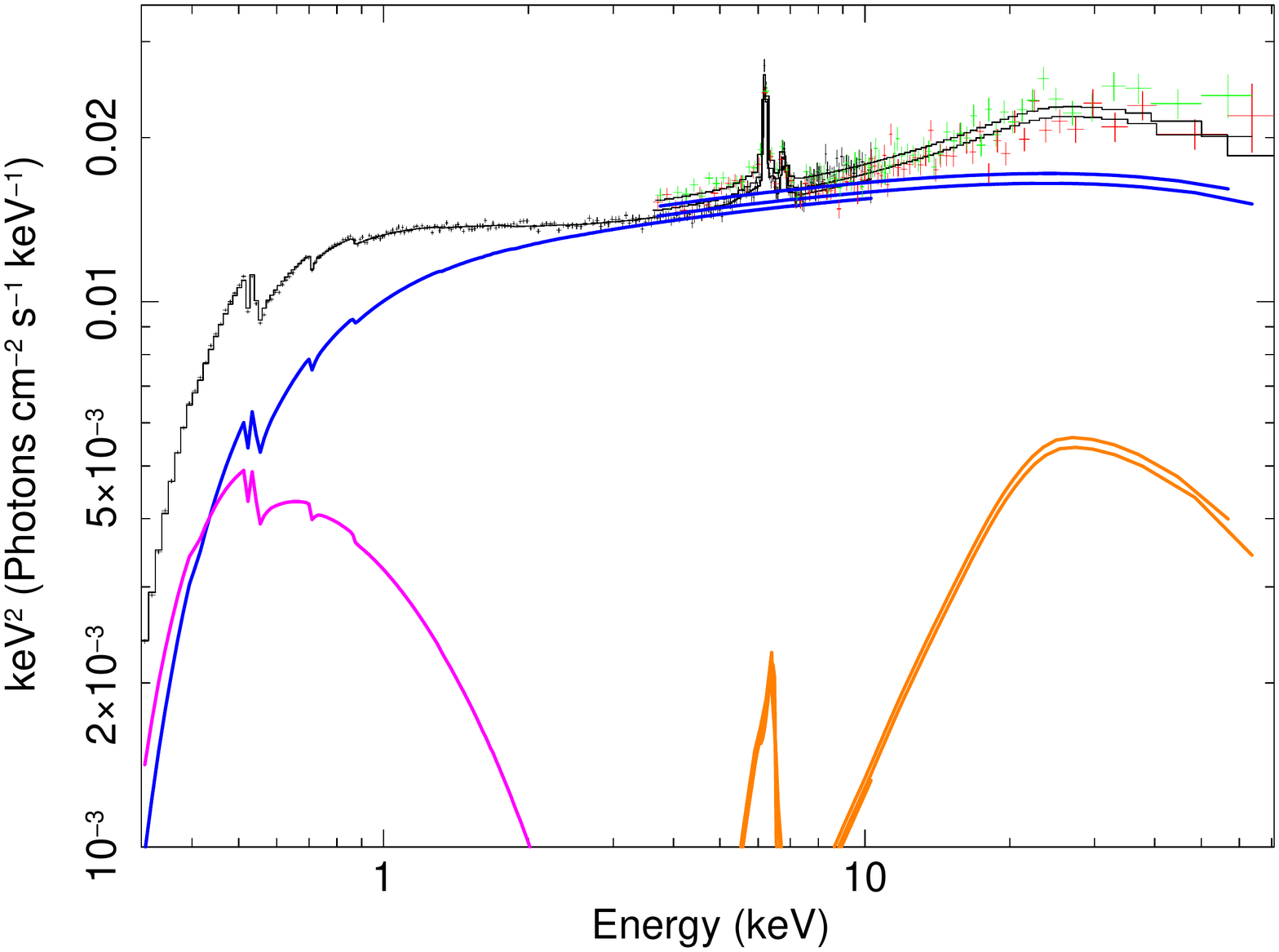} 
\end{tabular}
\caption{Fit over the 0.3--79\,keV energy range 
of the 2014 March 22 {\sl XMM-Newton}/pn and {\sl NuSTAR} data with
model $\mathcal{C}$, combining a soft Comptonization ({\sl
  comptt}), a cut-off powerlaw, and a relativistic reflection
component {\sc relxill} (as produced by 
the cut-off power-law primary continuum shape). 
The corresponding fit parameters are reported in
Table~\ref{tab:comptt}.
Black: {\sl XMM-Newton/pn} spectrum, red: {\sl NuSTAR} FPMA, and green:
{\sl NuSTAR} FPMB. 
{\it Top panel:} data and data/model ratio. 
{\sl Bottom panel:} Unfolded spectra where the contribution of the model components are
displayed. The following colour code for the emission components 
(continuous lines) is used: magenta 
 for {\sc comptt} (soft Comptonization), blue for the {\sc
   cut-off power law} (hot Comptonization), 
 orange for the relativistic reflection ({\sc relxill}), and black for
 the total emission. 
For clarity purpose we have not displayed the total model
and the three BLR Gaussian lines. }  
\label{fig:pnnustarmodelC}
\end{figure}

The cut-off power-law component is used here in order to merely
parametrize Comptonization from hot electrons of the thin corona. 
This continuum shape is also the one used as underlying hard X-ray
continuum for the relativistic reflection component.  
 Since the mildly relativistic Fe\,K component(s) do not appear to be 
 formed in the very inner part of the accretion disk, as found previously in this 
 work (see also Paper\,II), we allow $R_{\rm in}$ to vary and fix the emissivity index to 3. 
 We find a good fit ($\chi^{2}$/d.o.f.=2197.6/2058, $\chi^{2}_{\rm red}$=1.07), as illustrated in 
Fig.~\ref{fig:pnnustarmodelC} (top panel). 
The inferred fit parameters are reported in Table~\ref{tab:comptt}. 
We notice that the excess found above about 8\,keV in the residuals for
the pn spectrum is likely due to calibration issue between pn spectrum
using the small window mode and the NuSTAR spectra, and is
particularly prominent for high S/N pn spectrum\footnote{We note 
  that this pn excess above 8\,keV (when fitting simultaneous {\sl
    NuSTAR} spectra) is present even with
  the latest SAS version (16.0.0) and calibrations available in April 2017.}. 
However, removing pn data above 8\,keV leads to compatible fit
parameters within their error bars (except for
$\Gamma$=1.91$^{+0.02}_{-0.01}$, the discrepancy, however, is
marginal), with $\chi^{2}$/d.o.f.=1972.9/1876 
($\chi^{2}_{\rm red}$=1.05).   

We confirm that the Comptonizing plasma responsible for the soft excess
has a low temperature
($kT_{\rm e}$$\sim$0.5\,keV) and a high optical depth
($\tau$$\sim$9).  
Another interesting result is that during this 2014 observation, 
the source spectrum above 3\,keV shows a softer power-law index
($\Gamma$$\sim$1.87) 
compared to the 2013 one ($\Gamma\sim$1.73; \citealt{Matt14}), 
in agreement with the recent study based on the {\sl Swift} monitoring of Ark\,120, 
 which suggests that the source has a {\it ``steeper when brighter''} behaviour 
\citep[][see also Paper III]{Gliozzi17}.
The value of $R_{\rm in}$ of 25.5$^{+40.9}_{-8.0}$\,$R_{\rm g}$ is
much larger than the innermost stable orbit even for a maximal retrograde spin 
of the black hole, i.e a=$-$0.998. 
This is another hint that the observed Fe\,K lines are not formed in the 
very inner part of the accretion disk. 

The relative contribution of the different model components is displayed in
Fig.~\ref{fig:pnnustarmodelC} (unfolded spectrum for illustration
purposes only; bottom panel), and shows 
that the Comptonization process (including both warm and hot electrons
of the corona) is the dominant one over the broadband X-ray range.

\begin{table}[t!]
\caption{Simultaneous fit of the 2014 March 22 {\sl XMM-Newton}/pn and {\sl
  NuSTAR} spectra over the 0.3--79\,keV energy range with  model
$\mathcal{C}$, which combines Comptonization and relativistic
reflection contributions. The disk emissivity index ($q$) is fixed to
the standard value of 3.0. 
{\it nc} means that the parameter value is unconstrained. }           
\centering                          
\begin{tabular}{@{}l c }       
\hline\hline                
Parameters & model\,$\mathcal{C}$ \\  
\hline                       
$N_{\rm H}$ ($\times$10$^{21}$\,cm$^{-2}$)    &  1.04$\pm$0.03\\      
$kT_{\rm e}$ (keV) & 0.48$^{+0.07}_{-0.05}$ \\
$\tau$ & 9.0$\pm$0.7 \\
$norm$   & 2.9$^{+0.4}_{-0.3}$ \\
$\Gamma$ & 1.87$\pm$0.02 \\
$E_{\rm cut}$ (keV) & 183$^{+83}_{-43}$   \\
$norm$ & 1.2$\pm$0.1$\times$10$^{-2}$ \\
$a$ & {\it nc} \\
$\theta$ (degrees)  & 30(f)  \\
$A_{\rm Fe}$ & 1(f)\\
$R_{\rm in}$ ($R_{\rm g}$) &  25.5$^{+40.6}_{-8.0}$ \\
log\,$\xi$  (erg\,cm\,s$^{-1}$)& $\leq$0.1   \\
$\mathcal{R}$  & $\geq$0.3    \\  
$norm$ & 1.9$^{+0.5}_{-0.4}$$\times$10$^{-4}$   \\
\hline
C$_{\rm NuSTAR\,A}$  & 1.028$\pm$0.009  \\
C$_{\rm NuSTAR\,B}$  & 1.071$\pm$0.009  \\
 \hline
 \hline
L$^{\rm (a)}$ (0.3--2\,keV) &1.4$\times$10$^{44}$ \\
L$^{\rm (a)}$ (2--10\,keV) &1.0$\times$10$^{44}$ \\
L$^{\rm (a)}$ (10--79\,keV)  & 1.6$\times$10$^{44}$ \\
 \hline
 \hline
$\chi^{2}$/d.o.f.& 2197.6/2058   \\
$\chi^{2}_{\rm red}$& 1.07\\
\hline    
\hline                  
\end{tabular}
\tablefoot{(a) Unabsorbed luminosities, expressed in units
of erg\,s$^{-1}$. }
\label{tab:comptt}
\end{table}


\section{Summary}\label{sec:summary}

This paper is the fourth of a series of articles reporting on the
study of an extensive X-ray (and also optical and UV) observational 
campaign of a bare AGN, targeting Ark\,120. 
Here, we first performed the simultaneous spectral analysis of the
four 120\,ks {\sl XMM-Newton}/pn spectra obtained in March 2014, and then we analysed the
broadband X-ray spectrum -- combining pn and
{\sl NuSTAR} spectra -- obtained on 2014 March 22. 
The main results are summarized below: \\
 
(i) The four {\sl XMM-Newton} observations of Ark 120 in 2014 reveal both a strong
and variable soft excess and complex Fe\,K$\alpha$ emission. Above 3 keV,
both the continuum (with $\Gamma$$\sim$1.9) and the iron-line complex are similar
 with limited variability between the {\sl XMM-Newton} sequences.\\ 

(ii) Above 3 keV, the four 2014 pn spectra can be well fitted with a
disk reflection ({\sc relxill}) model with moderate reflection fraction
($\mathcal{R}$$\sim$0.4--0.5), which requires a flat emissivity
profile or a large height of the X-ray source above the disk. Both of
these interpretations suggest that the disk reflection emission originates
much further away than the ISCO, at typical radii of tens of $R_{\rm g}$.\\

(iii) The extrapolation of the above reflection models leaves strong residuals
due to the soft excess below 3 keV. To account for this, the 
reflection models tend to extreme, finely tuned values requiring a high
degree of blurring and nearly maximal black hole spin, plus a steep ($\Gamma$$\sim$2.4)
continuum. However, such models produce a very smooth broadband spectrum, and do
not account for the red/blue wings of the Fe\,K$\alpha$ line, which 
require a flat emissivity index (or large $R_{\rm in}$) and a harder
continuum ($\Gamma$$\sim$1.9). \\

(iv) Likewise, when a reflection dominated model is applied to the
2014 {\sl NuSTAR} data above 10 keV, simultaneous with the third {\sl
  XMM-Newton} observation,  
it cannot fit the spectrum in the highest energy range, leaving a large excess of 
residual emission above 30 keV. Hence, 
reflection-only models from a constant-density, geometrically flat accretion disk
cannot simultaneously account for the soft excess, iron line and hard
excess in the broadband 0.3--79\,keV spectrum. \\

(v) Instead, the X-ray broadband spectrum can be readily accounted for by a
Comptonization model, whereby the soft and hard continuum components
arise from a two temperature (warm, hot) disk corona. The warm part of
the corona ($kT_{\rm e}$$\sim$0.5\,keV) produces the low energy part of the
soft X-ray excess and is optically thick ($\tau$$\sim$9).
A disk reflection component is still required, but it is less strong and
originates at tens of gravitational radii from the ISCO of the black
hole.

\section{Discussion and conclusion}\label{sec:discussion}

During this extensive X-ray observational campaign of the bare AGN Ark\,120, 
carried out in March 2014, the source was caught in a high flux state similar to the
2003 {\sl XMM-Newton} observation \citep{Vaughan04}, and about as
bright as the 2013 low-flux observation \citep{Matt14}. 
Based on the long-term {\sl Swift} monitoring of Ark\,120 \citep[][Paper\,III]{Gliozzi17}, 
this large {\sl XMM-Newton} program and the 2013 observation cover the 
typical high- to low-flux range observed in this source. 
Our spectral analysis
confirms the {\it ``softer when brighter''} behaviour of Ark\,120
\citep{Gliozzi17}, as commonly found in AGN
\citep[e.g.,][]{Markowitz03,P04a,Sobolewska09,Soldi14,Connolly16,Ursini16} 
and black-hole binaries systems
\citep[e.g.,][]{Remillard06,Done07,Wu08,Dong14} with accretion rates 
above 0.01. \\

From the analysis of the four {\sl XMM-Newton}/pn spectra (2014 March
18--24) and the March 22 {\sl XMM-Newton}
and {\sl NuSTAR} observations, we find that relativistic reflection models from 
a constant-density, geometrically flat accretion disk  cannot self-consistently
reproduce the soft excess, the mildly relativistic red and
blue Fe\,K$\alpha$  features, and the hard X-ray excess.
We note that this shortcoming with relativistic
reflection models is found in all four 120\,ks {\sl XMM-Newton}/pn spectra
separately, thanks to the high-S/N spectra that can be obtained for a bright 
source like Ark\,120.  
In order to form the large, smooth soft X-ray excess, extreme and tuned 
parameters are found for the ``pure'' relativistic reflection
scenario: a maximally rotating 
black hole ($a$$\sim$0.998), a 
very centrally peaked disk emissivity, a soft primary photon index
($\Gamma$$\sim$2.4), and a very large reflection fraction
($\mathcal{R}$$\sim$10), as reported in Table~\ref{tab:pn}. 
By contrast, the red/blue Fe\,K$\alpha$ features require just a moderate 
reflection fraction ($\mathcal{R}$$\sim$0.3), a flat emissivity index 
($q$$\sim$1.6) or large $R_{\rm in}$ (a few tens of $R_{g}$), and a harder 
power-law index ($\Gamma$$\sim$1.9). 
The former case (reflection-dominated soft X-ray excess) would correspond 
to a compact corona located very close to the black hole, while the
latter case (disk origin of the Fe\,K$\alpha$ features) would correspond to an extended 
corona or a lamppost geometry with a large height of the X-ray source above the disk. 
Both conditions cannot be therefore explained
 
Instead, the whole 0.3--79\,keV spectrum can be readily explained by a
combination of Comptonization (dominating process) of the
thermal optical-UV seed photons from the accretion disk by a warm
($kT_{\rm e}$$\sim$0.5\,keV) optically thick plasma ($\tau$$\sim$9) below
about 0.5\,keV, by a hot optically thin corona above 0.5\,keV, 
 and mildly relativistic reflection at a few $\times$10\,$R_{\rm g}$.  
As shown in \cite{Rozanska15}, such a high optical
depth of the warm corona could mean that either a strong magnetic
field or vertical outflows to stabilize the system are required.
The in-depth investigation of 
the physical properties of the warm and hot corona will be performed in a
forthcoming paper (Tortosa et al.\, in preparation).  
Interestingly, such a soft X-ray excess origin (i.e., enabling to rule out a
relativistic reflection scenario too) is similar to that found, for example, in
 some non-bare AGN  
like \object{Mrk\,509} \citep{Mehdipour11,Petrucci13,Boissay14} and 
\object{NGC\,5548} \citep{Mehdipour15}, from deep X-ray (and multi-wavelength) 
observational campaigns. Such an origin has also been found for the lower-mass SMBH
AGN \object{NGC\,4593}, from  high-energy monitoring with {\sl
  XMM-Newton} and {\sl NuSTAR} \citep{Ursini16}. \\

Contrary to the 2013 observation, a relativistic reflection component is still required to
explain part of the Fe\,K$\alpha$ complex. 
This could be the signature that the optically
thick corona is hiding partly or totally the inner accretion disk, as
proposed by \cite{Matt14} to explain the X-ray characteristics of the
2013 observation, which is a factor of 2 lower in hard X-ray flux and
does not appear to show a broad Fe\,K$\alpha$ line component. 
In March 2014, the optically thick part of the corona may
have been less extended and/or have displayed a lower covering factor, 
allowing us to detect a larger part of the relativistic reflection emission. 
For example, as demonstrated by \cite{Wilkins15}, for a covering factor below about
85\%, the blurred reflection features become more detectable. Moreover, a 
Comptonizing corona that covers the inner regions of the accretion disk can
have substantial impact on the observed reflection spectrum
\citep{Petrucci01, Wilkins15, Steiner17}. 
Furthermore, if the corona covers a sufficient fraction of the inner
accretion disk so as to Comptonize the reflected emission, a low
reflection fraction can be measured, as found during these March 2014
observations ($\mathcal{R}$$\sim$0.3). This could support the presence
of an extended corona in this object.   
The case of a receding, full covering corona between February
2013 and March 2014 will be investigated through the fitting of the
spectral energy distribution from optical/UV to hard-X-rays in a
forthcoming paper (Porquet et al., in preparation). The alternative scenario
of a patchy corona will be tested in a future work too (Wilkins et
al.\, in preparation). \\
 
In conclusion, the great advantage of a source like Ark\,120 is that 
its ``bare'' properties remove any
fit degeneracy with warm absorption contributions. Thanks to both 
{\sl XMM-Newton} (4$\times$$\sim$120\,ks) and
{\sl NuSTAR} ($\sim$65\,ks), we are able to discriminate between Comptonization
and relativistic reflection for the soft X-ray excess origin, as
well as emission above about 2\,keV.  
As revealed in paper\,I thanks to
the very deep 2014 RGS spectrum, a substantial amount of X-ray
emitting warm gas is present out of the direct line of sight.
 The presence of this warm gas (warm absorber) is seen in a very large
amount of type I AGN \cite[e.g.,][]{P04a,Piconcelli05,Blustin05} via
mainly absorption lines, but also via emission lines
\citep[e.g.,][]{Bianchi02,Nucita10,WangJ11,Ebrero11,Reeves13}.  
 Such result demonstrates that Ark\,120 is not a peculiar 
source but merely a source where the line-of-sight does not intercept
the warm gas, and then broadly fits into the AGN unified scheme. 
Therefore, since Ark\,120 has typical AGN properties, such as mass and 
accretion rate, it can be used as a prototype to perform an in-depth
study of the X-ray corona and of its physical and geometrical properties over time, 
and of its possible impact on reflection spectra.\\

\begin{acknowledgements}
 The authors would like to deeply thank the anonymous referee for useful
and constructive comments.  
Based on observations obtained with the {\sl XMM-Newton}, and ESA science
mission with instruments and contributions directly funded by ESA
member states and the USA (NASA). 
 This work made use of data from the
{\sl NuSTAR} mission, a project led by the California Institute of
Technology, managed by the Jet Propulsion Laboratory, and
funded by NASA. 
This research has made use of the {\sl NuSTAR} 
Data Analysis Software (NuSTARDAS) jointly developed by
the ASI Science Data Center and the California Institute of
Technology.
D.P.\ would like to acknowledge financial support from the French ``Programme
National Hautes Energies'' (PNHE). Part of the work was supported by
the European Union Seventh Framework Program under the grant
agreement No.\ 312789 (DP, GM, AM, AF).  
JNR acknowledges financial support via Chandra grant number GO4-15092X
and NASA grant NNX15AF12G. JNR and AL also acknowledge support of the
STFC, via the consolidated grant ST/M001040/1. 
GM, AM, AT and FU acknowledge financial support from Italian Space
Agency under grant ASI/INAF I/037/12/0-011/13. 
E.N.\ acknowledges funding from the European Union's Horizon 2020
research and innovation programme under the Marie
Sklodowska-Curie grant agreement No. 664931. 
\end{acknowledgements}

%
%


\appendix
\section{Description of the relativistic reflection package relxill} \label{ref:relxill}

\subsection{Coronal geometry}
The {\sc relxill} models are characterized by the
following parameters: \\
$-$ the photon index of the illuminating radiation (identical for both the intrinsic
cut-off power law and the relativistic reflection spectrum): $\Gamma$;\\
$-$ the black hole spin: $a$;\\
$-$ the disk inclination angle: $\theta$;\\
$-$ the inner and outer radii of the disk: $R_{\rm in}$ and $R_{\rm out}$,
respectively;\\
$-$ the broken power-law disk emissivity index: $q_1$ (for $R<R_{\rm br}$), $q_2$
(for $R$$>$$R_{\rm br}$) and  the radius ($R_{\rm br}$) where emissivity changes from $q_1$ to
$q_2$;\\ 
$-$ the reflection fraction as defined in \cite{Dauser16}: $\mathcal{R}$; \\
$-$ the ionization parameter (erg\,cm\,s$^{-1}$, in log units) at the
surface of the disk 
(i.e., the ratio of the X-ray flux to the gas density): log\,$\xi$; \\ 
$-$ the iron abundance relative to the solar value (here \citealt{Grevesse98}):
$A_{\rm Fe}$; \\
$-$ the high-energy cut-off (identical for both the intrinsic
power law and the relativistic reflection spectrum): $E_{\rm cut}$. 

The {\sc relxill$\_$ion} model is similar to the {\sc relxill} one 
but allows us to calculate the reflection from the disk with
several zones of different ionization (log\,$\xi$ $\propto$ $R^{\rm -index}$). 
The number of the zones (which has to be high enough for a good
physical  representation, here taken at 15) and the ionization gradient versus the radius
can be specified directly in the model.

\subsection{Lamppost geometry}

The {\sc relxill$\_$lp} models are defined for the lamppost geometry. 
The parameters $q_1$, $q_2$, and $R_{\rm br}$ are replaced  by the height of the
primary source $h$, and the reflection fraction value is
self-consistently determined by the the lamppost geometry ({\sc fixReflFrac=1}).

The {\sc relxill$\_$lp$\_$ion} model takes into account 
an ionization gradient of the accretion disk as for the {\sc
  relxill$\_$ion} model, 
while the {\sc relxill$\_$lp$\_$alpha} model calculates self-consistently the radial 
dependence of the ionization from the
irradiation of the disk, using a certain mass accretion rate
($\dot{m}$) 
and assuming the density profile of an $\alpha$ accretion disk \citep{Shakura73}.

\section{Reflection models for the four 2014 XMM-Newton/pn spectra}

\begin{table*}[t!]
\caption{Simultaneous fit of the four 2014 {\sl XMM-Newton}/pn spectra with the
  baseline relativistic reflection  model (model $\mathcal{A}$, $n$$=$10$^{15}$\,cm$^{-3}$) over the 0.3--10\,keV energy range. }             
\centering                          
\begin{tabular}{lccccccccc}        
\hline\hline                 
Parameters & \multicolumn{1}{c}{relxill} &
\multicolumn{1}{c}{relxill$\_$tbp}  &
\multicolumn{1}{c}{relxill$\_$ion}  & \multicolumn{1}{c}{relxilllp}&
\multicolumn{1}{c}{relxilllp$\_$alpha}&
\multicolumn{1}{c}{relxilllp$\_$ion}   \\    
\hline                       
$N_{\rm H}$ ($\times$10$^{21}$\,cm$^{-2}$)  &  1.09$\pm$0.02  &  1.06$\pm$0.01  &  1.18$\pm$0.02  & 1.00$\pm$0.01 &  1.15$\pm$0.01 & 1.13$\pm$0.01 \\      
$a$         & 0.965$^{+0.003}_{-0.002}$ &  0.970$\pm$0.001  &
$\geq$0.997 & 0.968$^{+0.003}_{-0.005}$   & $\geq$0.992  & $\geq$0.997\\
$\theta$ (degrees)&$\leq$6.7 & $\leq$6.3 &  30.6$\pm$0.9  &$\leq$5.9 & $\leq$5.6& $\leq$5.9 \\
log\,$\xi$  (erg\,cm\,s$^{-1}$)        & $\leq$0.1 & $\leq$0.1 & $\geq$4.5  & $\leq$0.1& $-$ &  3.8$\pm$0.1 \\
$ion\_index$      & $-$ & $-$ & 7.9$^{+0.4}_{-0.6}$ & $-$& $-$& $-$ \\
$\dot{m}$ (\%)& $-$ & $-$ & $-$ & $-$& 18.2$\pm$0.3 & 18.2(f) \\
$density\_index$  & $-$ & $-$ & $-$ & $-$& $-$ & 1.6$^{+0.3}_{-0.2}$ \\
$A_{\rm Fe}$      & $\leq$0.5 & $\leq$0.5  & 0.8$\pm$0.1  & $\leq$0.5 &$\leq$0.5 & $\leq$0.5\\
\hline                                  
                          &   \multicolumn{5}{c}{2014 March 18} \\
\hline                                  
$\Gamma$  & 2.48$\pm$0.02 &  2.46$\pm$0.01  & 2.38$\pm$0.01 & 2.37$\pm$0.01& 2.40$\pm$0.01& 2.39$\pm$0.01\\
$\mathcal{R}$       &  10.8$^{+1.1}_{-0.9}$ & 10.2$\pm$0.1  & 15.2$^{+2.0}_{-1.8}$ & $-$& $-$& $-$\\
$norm$  & 2.0$\pm0.1\times$10$^{-4}$ &2.0$\pm0.1\times$10$^{-4}$ & 1.4$\pm0.1\times$10$^{-4}$  & 1.5$\pm0.1\times$10$^{-2}$ & 1.9$\pm0.1\times$10$^{-2}$& 1.7$\pm$0.1$\times$10$^{-2}$ \\
$q_1$ &  7.4$^{+0.6}_{-0.5}$ & 6.5$\pm$0.1   &  7.6$\pm$0.4 & $-$& $-$& $-$ \\
$q_2$ & 4.8$\pm$0.5  & $-$2.5$^{+0.6}_{-0.8}$  & 3.1$\pm$0.4 & $-$& $-$& $-$\\
$q_3$ & $-$ & 4.0$^{+0.2}_{-0.1}$ & $-$ & $-$ & $-$ & $-$ \\
$R_{\rm br1}$ ($R_{\rm g}$) & 3.8$^{+1.0}_{-0.5}$ & 6.3$\pm$0.1  & 3.9$^{+0.7}_{-0.5}$  & $-$& $-$& $-$\\
$R_{\rm br2}$ ($R_{\rm g}$) & $-$ & $\leq$8.3 &$-$ &$-$
&$-$& $-$\\
$h$ ($R_{\rm g}$)   & $-$ & $-$ & $-$  & 1.8$\pm$0.1 & 1.7$\pm$0.1  & 1.7$\pm$0.1  \\
\hline                                  
                &   \multicolumn{5}{c}{2014 March 20} \\
\hline                                  
$\Gamma$      & 2.37$\pm$0.02 & 2.36$\pm$0.01   & 2.30$\pm$0.01 & 2.29$\pm$0.01& 2.33$\pm$0.01& 2.31$\pm$0.01\\
$\mathcal{R}$            & 7.5$\pm$0.7  & 7.4$\pm$0.1   &  11.3$^{+1.8}_{-1.5}$  & $-$& $-$& $-$\\
$norm$       & 1.7$\pm0.1\times$10$^{-4}$  &  1.7$\pm0.1\times$10$^{-4}$     & 1.3$\pm0.1\times$10$^{-4}$ & 1.0$\pm0.1\times$10$^{-2}$ 
& 1.3$\pm0.1\times$10$^{-2}$& 1.2$\pm0.1\times$10$^{-2}$   \\
$q_1$          &6.8$^{+0.7}_{-0.6}$     & 6.1$\pm$0.1  &  7.5$\pm$0.6  & $-$& $-$& $-$ \\
$q_2$         & 3.4$^{+0.4}_{-0.3}$  & $-$1.1$^{+0.6}_{-0.8}$  & 2.9$\pm$0.3 & $-$& $-$& $-$\\
$q_3$          & $-$ & 3.5$^{+0.8}_{-0.2}$ &$-$ &$-$ &$-$ & $-$ \\
$R_{\rm br1}$ ($R_{\rm g}$) & 5.2$^{+1.3}_{-1.0}$ & 9.4$^{+0.3}_{-0.1}$  & 3.7$^{+0.6}_{-0.5}$  & $-$& $-$& $-$\\
$R_{\rm br2}$ ($R_{\rm g}$) & $-$ & 17.1$^{+1.1}_{-0.6}$ & $-$ &$-$
&$-$& $-$\\
$h$ ($R_{\rm g}$)    & $-$ & $-$ & $-$  & 1.9$\pm$0.1 & 1.7$\pm$0.1 & 1.7$\pm$0.1   \\
\hline                                 
                &   \multicolumn{6}{c}{2014 March 22} \\
\hline                                  
$\Gamma$  &  2.38$\pm$0.02 & 2.36$\pm$0.01  & 2.29$\pm$0.01 & 2.27$\pm$0.01& 2.31$\pm$0.01& 2.30$\pm$0.01 \\
$\mathcal{R}$    & 8.3$^{-0.9}_{-0.8}$    & 8.1$\pm$0.1  & 12.8$^{-1.9}_{-1.8}$ &$-$&$-$&$-$\\
$norm$    & 1.9$\pm0.1\times$10$^{-4}$ & 1.9$\pm0.1\times$10$^{-4}$  &1.3$\pm$0.1$\times$10$^{-4}$  & 1.1$\pm0.1\times$10$^{-2}$ & 1.3$\pm0.1\times$10$^{-2}$ & 1.3$\pm0.1\times$10$^{-2}$ \\
$q_1$    &7.6$^{+0.8}_{-0.5}$   & 6.7$\pm$0.1  &  7.5$\pm$0.5 & $-$& $-$& $-$ \\
$q_2$    & 5.0$\pm$0.5  & 1.3$\pm$0.5  & 3.2$\pm$0.5 & $-$& $-$& $-$\\
$q_3$     & $-$ & $\leq$7.7 &$-$ &$-$ &$-$ & $-$ \\
$R_{\rm br1}$ ($R_{\rm g}$) & 3.4$^{+0.8}_{-0.5}$ & 6.1$^{+0.3}_{-0.1}$  & 3.8$^{+0.9}_{-0.5}$  & $-$& $-$& $-$\\
$R_{\rm br2}$ ($R_{\rm g}$) & $-$ & 9.4$^{+2.5}_{-0.2}$ & $-$ &$-$
&$-$& $-$\\
$h$ ($R_{\rm g}$)   & $-$ & $-$ & $-$ & 1.9$\pm$0.1 & 1.7$\pm$0.1 & 1.7$\pm$0.1   \\
 \hline
                &   \multicolumn{5}{c}{2014 March 24} \\
\hline                                  
$\Gamma$ & 2.37$\pm$0.02  & 2.36$\pm$0.01  &2.29$\pm$0.01  & 2.27$\pm$0.01 & 2.31$\pm$0.1& 2.30$\pm$0.01 \\
$\mathcal{R}$    & 8.2$^{+0.8}_{-0.7}$   &  8.2$^{+0.5}_{-0.1}$  & 13.6$^{+2.5}_{-2.0}$    & $-$& $-$& $-$\\
$norm$ & 1.7$\pm0.1\times$10$^{-4}$ &  1.7$\pm0.1\times$10$^{-4}$   &  1.2$\pm0.1\times$10$^{-2}$  & 1.0$\pm0.1\times$10$^{-2}$ & 1.3$\pm0.1\times$10$^{-2}$ & 1.2$\pm0.1\times$10$^{-2}$ \\
$q_1$ &  8.1$^{+0.7}_{-1.0}$   & 6.8$^{+0.2}_{-0.1}$   &  8.1$^{+0.6}_{-0.5}$ & $-$& $-$& $-$ \\
$q_2$  & 4.3$^{+0.5}_{-0.6}$  &$-$3.1$^{+0.7}_{-0.6}$   & 3.3$\pm$0.3 & $-$& $-$& $-$\\
$q_3$  & $-$ & $\leq$3.6  &$-$ &$-$ &$-$ & $-$\\
$R_{\rm br1}$ ($R_{\rm g}$) & 3.6$^{+1.7}_{-0.5}$ &  8.0$\pm$0.4  & 3.2$\pm$0.4  & $-$& $-$& $-$\\
$R_{\rm br2}$ ($R_{\rm g}$) & $-$ & 13.2$^{+0.2}_{-0.1}$ &$-$ &$-$ &$-$ & $-$\\
$h$ ($R_{\rm g}$)    & $-$ & $-$  &$-$  & 1.8$\pm$0.1 & 1.7$\pm$0.1 & 1.7$\pm$0.1   \\
 \hline
 \hline
$\chi^{2}$/d.o.f.  &  7246.6/6728 & 7278.0/6720  & 7164.5/6727 & 7507.4/6740  &  7329.7/6740&  7309.1/6739 \\
$\chi^{2}_{\rm red}$ & 1.08  & 1.08   &1.07  &  1.11 & 1.09& 1.08\\
\hline    \hline                  
\end{tabular}
\label{tab:pn}
\end{table*}

Table~\ref{tab:pn} reports the spectral fits of the four 2014
{\sl XMM-Newton}/pn spectra using the {\sc relxill} models (see
$\ref{sec:allpnref})$.

\section{Reflection modelling for the X-ray broad-band spectra in
  2014 March 22}

\subsection{Investigation of possible difference between the observed
  and incident photon indices}\label{sec:complex}

Following the results from \cite{Fuerst15} for
the X-ray binary \object{GX 339-4} (in low-luminosity, hard states),
we then investigate if the inability of the relativistic reflection
models to account for the broad-band spectrum 
could be explained by a moderate difference in the spectral powerlaw
indices. For GX\,339--4 a $\Delta\Gamma$$\sim$0.3 were found\footnote{However,
we note that \cite{Steiner17} proposed that this hardening 
of reflection spectrum can be explained by Compton scattering in the corona.}. 

For the coronal geometry, we allow the blurred reflector to see a different continuum
  ($\Gamma$) with respect to 
  the primary one. Since in the {\sc relxill} model the intrinsic 
  continuum is included in the model, not enabling us to have a different
  continuum shape for the direct and the reflected one, 
 we modify the baseline model $\mathcal{A}$ as {\sc tbnew$\times$(cutoffpl + relxill
  + 3 zgaussians)}, with the reflection fraction of the {\sc
    relxill} model set to negative values to allow only the reflection
  component to be returned. 
{\sc cutoffpl} is a power law with a high-energy exponential
  cut-off. 
  The fit statistic improves significantly with $\Delta\chi^{2}$$\sim$$-$287 for only one additional
  parameter. However, this would mean an unlikely scenario 
with very different continua with
  $\Delta\Gamma$$\sim$1.0. 
  We then investigate a scenario with two
relativistic reflection models with different continuum shapes. 
This improves significantly the fit statistic compared to the single
reflection component, with $\Delta\chi^2$$\sim$176 for three 
additional parameters. We find $\Delta\Gamma$$\sim$0.4, but a
hard X-ray excess residual is still present above 35\,keV. 

We perform the same tests as above assuming a lamppost geometry, 
but we are unable to find satisfactory fits, even considering 
a vertically extended primary continuum source on the rotation
    axis ({\sc
      relxill$\_$lp$\_$ext$\otimes$xillver}) or a moving continuum
    source 
({\sc relxilllp$\_$velo}).

Of course in all above fits, the Fe\,K$\alpha$ red and blue residuals
are still present.

\subsection{High-density reflection modelling}\label{sec:highdens}

\begin{figure}[t!]
\begin{tabular}{c}
\includegraphics[width=0.9\columnwidth,angle=0]{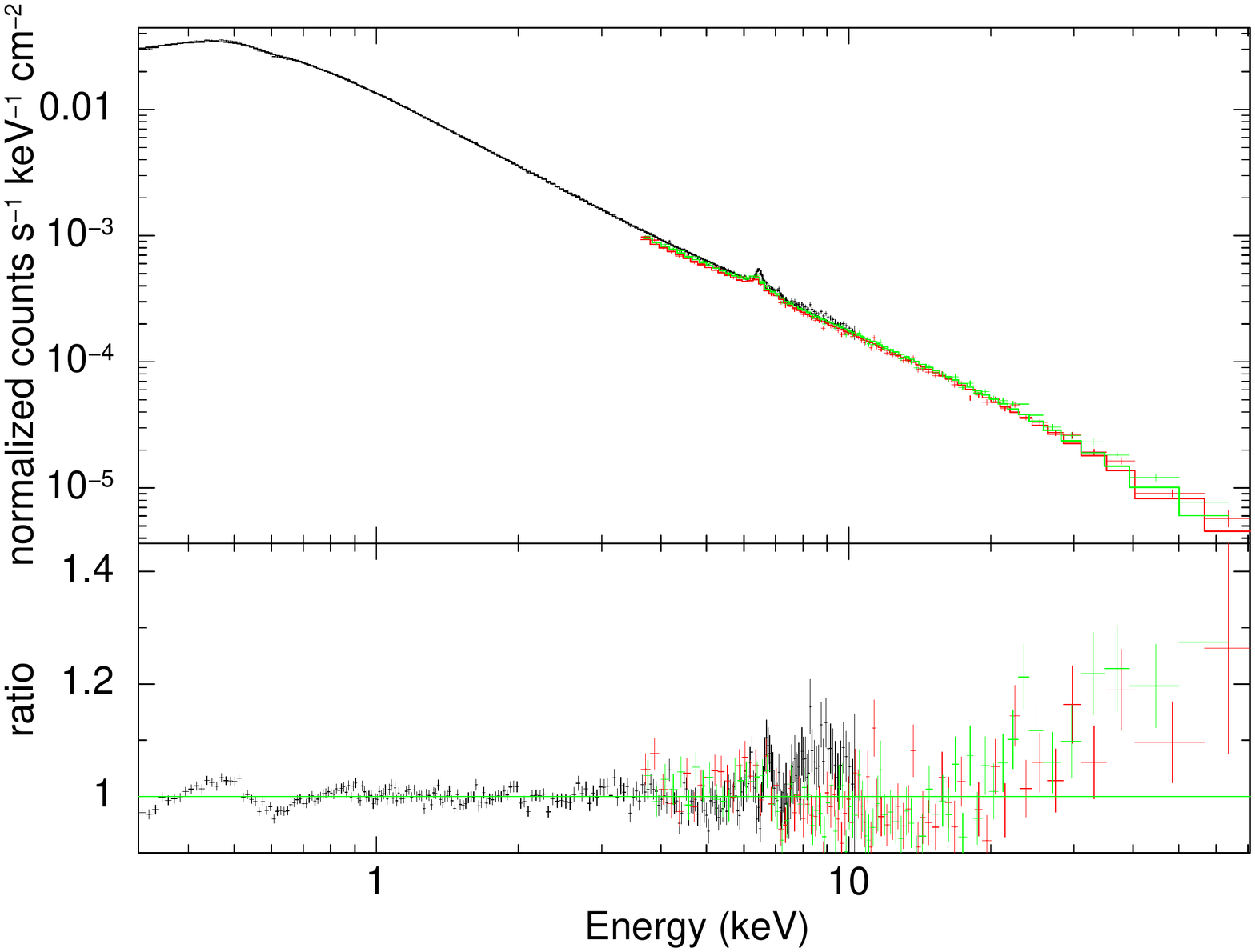}  
\end{tabular}
\caption{Fit over the 0.3--79\,keV energy range 
of the 2014 March 22 {\sl XMM-Newton/pn} and {\sl NuSTAR} data with a relativistic
reflection baseline model using {\sc relxill} (model $\mathcal{A}$,
$n$$=$10$^{19}$\,cm$^{-3}$). 
}  
\label{fig:pnnustarrelxill19}
\end{figure}

The reflection models used in this work are calculated
for a density of 10$^{15}$\,cm$^{-3}$, for which it is assumed that the
ionization state of the gas is identical for a given $\xi$ value. 
But, as reported very recently by \cite{Garcia16}, higher densities
{\it ``are most relevant for low mass accreting black holes and when the
coronal fraction is high''}. Therefore, they computed the reflected
spectra for densities larger than the commonly assumed density
of 10$^{15}$\,cm$^{-3}$, i.e., up to $n$=10$^{19}$\,cm$^{-3}$.
 They showed that, for such higher density models, 
a very significant increase in the continuum flux
at energies below about 2\,keV occurs due to a large increase of
thermal emission at soft X-ray energies. In other words, a
high-density accretion disk leads to a larger soft excess compared to 
lower-density disks (such as 10$^{15}$\,cm$^{-3}$). 
 Therefore, since such a high density may be appropriate for Ark\,120, 
a full grid of reflection models for $n$=10$^{19}$\,cm$^{-3}$ 
with high-energy cut-off fixed at 1000\,keV has been produced 
(J.\ Garcia, private communication). 
 We apply this high-density model to the 2014 March 22 {\sl XMM-Newton}/pn and {\sl NuSTAR} spectra,
for a coronal geometry. 
 We find an unsatisfactory fit of the X-ray broad-band energy
  ($\chi^2$/d.o.f.=2513.3/2058; $\chi_{\rm red}$=1.22). 
 A weaker -- though still significant -- hard X-ray excess is found
 since a harder photon powerlaw index ($\Gamma$=1.76$\pm$0.01) can produce a larger soft
 excess emission, while a soft
 photon index is required for lower disk density
 ($\Gamma$=2.22$\pm$0.01, see Table~\ref{tab:2014cref}).  
In addition, there are still significant model/data deviations below
  about 1\,keV and in the Fe\,K$\alpha$ energy range (see Fig.~\ref{fig:pnnustarrelxill19}). 
In conclusion, even a high-density disk cannot reproduce the
broad-band X-ray spectrum. 

\end{document}